\documentclass[11pt,a4paper]{article}
\usepackage[margin=2.5cm]{geometry}
\usepackage[T1]{fontenc}
\usepackage{lmodern}
\usepackage{textcomp}
\usepackage{amsmath,amssymb}

  \usepackage{setspace}
\usepackage[mathlines]{lineno}
\usepackage[super,sort&compress,comma]{natbib}
\usepackage{xurl}
\usepackage[hidelinks]{hyperref}
\hypersetup{pdfauthor={},pdfsubject={},pdfkeywords={}}

  \pdftrailerid{}

\usepackage{xcolor}
\usepackage{graphicx}
\graphicspath{{figures/}}

\newcommand{\setval}[2]{\expandafter\def\csname val@#1\endcsname{#2}}
\newcommand{\val}[1]{\ifcsname val@#1\endcsname\csname val@#1\endcsname\else\errmessage{Unknown ledger value #1}\fi}

\newcommand{\figlegend}[2]{\par\bigskip\noindent\textbf{#1\,\textbar\,#2}\ }
\newcommand{\pnl}[1]{\textbf{#1}}
\newcommand{\setlegend}[2]{\expandafter\def\csname leg@#1\endcsname{#2}}
\newcommand{\uselegend}[1]{\ifcsname leg@#1\endcsname\csname leg@#1\endcsname\else\errmessage{No legend for #1}\fi}

\newcommand{\Pv}{\ensuremath{P}}
\newcommand{\ci}[2]{95\%~CI #1 to #2}

\setval{B1}{0.376}%
\setval{B2}{+0.013}%
\setval{B3}{0.004}%
\setval{B4}{0.022}%
\setval{B5}{0.016}%
\setval{B6}{+0.036}%
\setval{B7}{0.385}%
\setval{B8}{+0.019}%
\setval{B9}{0.014}%
\setval{B10}{0.031}%
\setval{B11}{0.341}%
\setval{B12}{+0.025}%
\setval{B13}{0.070}%
\setval{B14}{0.389}%
\setval{B15}{0.264}%
\setval{B16}{0.256}%
\setval{B17}{1,726}%
\setval{B18}{1,668}%
\setval{B19}{37.5}%
\setval{B20}{96.6}%
\setval{B21}{2.6}%
\setval{B22}{2.3}%
\setval{B23}{0.081}%
\setval{B24}{52\%}%
\setval{B25}{23\%}%
\setval{B26}{45\%}%
\setval{B27}{+0.011}%
\setval{B28}{0.011}%
\setval{B29}{0.028}%
\setval{B30}{0.006}%
\setval{B31}{0.044}%
\setval{B32}{+0.018}%
\setval{B33}{0.010}%
\setval{B34}{0.354}%
\setval{B35}{0.369}%
\setval{B36}{0.437}%
\setval{B37}{0.417}%
\setval{B38}{16\%}%
\setval{B39}{11\%}%
\setval{B40}{57\%}%
\setval{B41}{0.069}%
\setval{B42}{1,691}%
\setval{B43}{1,764}%
\setval{B45}{0.186}%
\setval{B46}{0.187}%
\setval{B47}{0.275}%
\setval{B48}{0.239}%
\setval{B49}{1,040}%
\setval{B50}{1,041}%
\setval{B51}{13}%
\setval{B52}{0.01}%
\setval{B53}{+0.027}%
\setval{D1}{1,160}%
\setval{D2}{983}%
\setval{D3}{163,989}%
\setval{D4}{174}%
\setval{D5}{148}%
\setval{D6}{175}%
\setval{D7}{44}%
\setval{D8}{194,265}%
\setval{D9}{29,851}%
\setval{D10}{107}%
\setval{D11}{113}%
\setval{D12}{62}%
\setval{D13}{13}%
\setval{D14}{581,503}%
\setval{D15}{20,010}%
\setval{D16}{6.0}%
\setval{D17}{219.6}%
\setval{D18}{219,595,334}%
\setval{D19}{921,272}%
\setval{D20}{3}%
\setval{D21}{1,157}%
\setval{D22}{193,840}%
\setval{D23}{6,041,065}%
\setval{D24}{164}%
\setval{H1}{+0.059}%
\setval{H2}{+0.013}%
\setval{H3}{+0.015}%
\setval{H4}{\textminus{}0.79}%
\setval{H5}{0.038}%
\setval{H6}{0.081}%
\setval{H7}{\textminus{}0.001}%
\setval{H8}{0.027}%
\setval{H9}{0.001}%
\setval{H10}{0.030}%
\setval{H11}{+0.035}%
\setval{H12}{0.017}%
\setval{H13}{0.054}%
\setval{H14}{+0.011}%
\setval{H15}{\textminus{}0.006}%
\setval{H16}{0.028}%
\setval{H17}{\textminus{}0.66}%
\setval{H18}{73}%
\setval{H19}{47}%
\setval{H20}{45}%
\setval{H21}{33}%
\setval{H23}{4}%
\setval{H25}{48}%
\setval{P1}{0.180}%
\setval{P2}{0.260}%
\setval{P3}{\textminus{}0.080}%
\setval{P4}{75\%}%
\setval{P5}{\textminus{}0.108}%
\setval{P6}{\textminus{}0.057}%
\setval{P7}{0.69}%
\setval{P8}{0.205}%
\setval{P9}{0.274}%
\setval{P10}{64}%
\setval{P12}{1}%
\setval{P13}{40}%
\setval{P14}{16}%
\setval{P15}{20}%
\setval{P16}{30}%
\setval{A1}{0.498}%
\setval{A2}{0.451}%
\setval{A3}{+0.045}%
\setval{A4}{0.022}%
\setval{A5}{0.070}%
\setval{A6}{+0.041}%
\setval{A7}{0.017}%
\setval{A8}{0.064}%
\setval{A9}{47}%
\setval{A10}{0.402}%
\setval{A11}{0.820}%
\setval{A12}{0.794}%
\setval{A13}{10\%}%
\setval{E1}{+0.851}%
\setval{E2}{0.711}%
\setval{E3}{0.991}%
\setval{E4}{104}%
\setval{E5}{52}%
\setval{E6}{+0.128}%
\setval{E7}{0.028}%
\setval{E8}{0.239}%
\setval{E9}{22}%
\setval{E10}{11}%
\setval{E11}{\textminus{}0.018}%
\setval{E12}{\textminus{}0.048}%
\setval{E13}{\textminus{}0.001}%
\setval{E14}{0.001}%
\setval{E15}{0.019}%
\setval{E16}{90}%
\setval{E17}{142}%
\setval{E18}{4.1}%
\setval{E19}{6.1}%
\setval{E20}{10.2}%
\setval{E21}{21}%
\setval{D25}{220}%
\setval{B54}{43}%
\setval{B55}{32}%
\setval{B56}{64}%
\setval{B57}{128}%
\setval{B58}{2,000}%
\setval{H26}{34}%
\setval{P17}{0.05}%
\setval{E22}{50}%
\setval{D26}{57}%
\setval{D27}{134}%
\setval{D28}{40}%
\setval{A14}{0.451}%
\setval{D29}{999}%
\setval{D30}{10,000}%
\setval{P18}{25}%
\setval{B63}{1,691}%
\setval{B64}{1,656}%
\setval{B65}{1,730}%
\setval{B66}{1,647}%
\setval{B67}{27}%
\setval{B68}{16}%
\setval{B69}{158}%
\setval{E23}{2.34}%
\setval{E24}{2.04}%
\setval{E25}{2.69}%
\setval{E26}{1.14}%
\setval{E27}{1.03}%
\setval{E28}{1.27}%
\setval{A15}{46}%
\setval{H28}{34}%
\setval{H29}{24}%
\setval{B70}{+0.014}%
\setval{B71}{\textminus{}0.008}%
\setval{B72}{0.037}%
\setval{B73}{0.21}%
\setval{B74}{12}%
\setval{B75}{+0.008}%
\setval{B76}{\textminus{}0.007}%
\setval{B77}{0.022}%
\setval{B78}{+0.001}%
\setval{B79}{\textminus{}0.025}%
\setval{B80}{0.026}%
\setval{B81}{\textminus{}0.0002}%
\setval{B82}{+0.008}%
\setval{B83}{0.005}%
\setval{B84}{0.029}%
\setval{B88}{0.071}%
\setval{B89}{0.062}%
\setval{B90}{5}%
\setval{B91}{4}%
\setval{B93}{5}%
\setval{C1}{137}%
\setval{C2}{0.382}%
\setval{C3}{+0.015}%
\setval{C4}{0.004}%
\setval{C5}{0.026}%
\setval{C6}{0.018}%
\setval{S1}{+0.010}%
\setval{S2}{0.002}%
\setval{S3}{0.018}%
\setval{S4}{0.054}%
\setval{S5}{+0.021}%
\setval{S6}{+0.011}%
\setval{S7}{0.002}%
\setval{S8}{0.021}%
\setval{S9}{0.059}%
\setval{S10}{+0.009}%
\setval{S11}{\textminus{}0.0004}%
\setval{S12}{0.018}%
\setval{S13}{0.189}%
\setval{S14}{+0.007}%
\setval{S15}{\textminus{}0.009}%
\setval{S16}{0.026}%
\setval{S17}{+0.012}%
\setval{S18}{+0.032}%
\setval{S19}{0.014}%
\setval{S20}{0.051}%
\setval{S21}{7}%
\setval{S22}{2}%
\setval{S23}{1}%
\setval{S24}{+0.025}%
\setval{S26}{+0.018}%
\setval{S28}{+0.020}%
\setval{S30}{0.452}%
\setval{S31}{0.443}%
\setval{S32}{0.430}%
\setval{S33}{0.281}%
\setval{S34}{0.280}%
\setval{S35}{0.257}%
\setval{S36}{7}%
\setval{G1}{0.111}%
\setval{G2}{49}%
\setval{G3}{34}%
\setval{G4}{64}%
\setval{G5}{+0.018}%
\setval{G6}{0.008}%
\setval{G7}{0.029}%
\setval{G8}{+0.007}%
\setval{G9}{\textminus{}0.007}%
\setval{G10}{0.021}%
\setval{G11}{0.39}%
\setval{G15}{\textminus{}0.023}%
\setval{G16}{\textminus{}0.043}%
\setval{G17}{\textminus{}0.008}%
\setval{G18}{\textminus{}0.012}%
\setval{G19}{0.378}%
\setval{G20}{0.370}%
\setval{H30}{+0.022}%
\setval{H31}{0.002}%
\setval{H32}{0.044}%
\setval{H33}{47}%
\setval{H34}{+0.020}%
\setval{H35}{0.003}%
\setval{H36}{0.037}%
\setval{H37}{+0.026}%
\setval{H38}{0.008}%
\setval{H39}{0.047}%
\setval{H40}{26}%
\setval{X6}{+0.015}%
\setval{X7}{0.009}%
\setval{X8}{0.021}%
\setval{X9}{+0.000}%
\setval{X10}{\textminus{}0.005}%
\setval{X11}{0.006}%
\setval{X12}{+0.017}%
\setval{X13}{0.012}%
\setval{X14}{0.022}%
\setval{X15}{\textminus{}0.002}%
\setval{X16}{\textminus{}0.007}%
\setval{X17}{0.002}%
\setval{V1}{+0.022}%
\setval{V2}{+0.033}%
\setval{V3}{+0.050}%
\setval{V4}{+0.028}%
\setval{V5}{+0.043}%
\setval{V6}{+0.040}%
\setval{K1}{34}%
\setval{K2}{27}%
\setval{K3}{7}%
\setval{K4}{38}%
\setval{K5}{15}%
\setval{K6}{23}%
\setval{K7}{143}%
\setval{K8}{204}%
\setval{K9}{4}%
\setval{K10}{7}%
\setval{K11}{6}%
\setval{K12}{10}%
\setval{K13}{19}%
\setval{K14}{14}%
\setval{K15}{30}%
\setval{S37}{6,000}%
\setval{X19}{+0.010}%
\setval{B94}{0.333}%
\setval{B95}{0.340}%
\setval{S38}{+0.018}%
\setval{S39}{0.005}%
\setval{S40}{0.029}%
\setval{B96}{+0.016}%
\setval{B97}{\textminus{}0.010}%
\setval{B98}{0.038}%
\setval{B99}{+0.002}%
\setval{B100}{\textminus{}0.020}%
\setval{B101}{0.021}%
\setval{B102}{+0.022}%
\setval{B103}{0.004}%
\setval{B104}{0.043}%
\setval{B105}{6}%
\setval{B106}{5}%
\setval{B107}{10}%
\setval{T1}{+0.013}%
\setval{T2}{\textminus{}0.001}%
\setval{T3}{0.027}%
\setval{T4}{+0.038}%
\setval{T5}{0.024}%
\setval{T6}{0.052}%
\setval{V7}{0.308}%
\setval{V8}{0.292}%
\setval{V9}{+0.041}%
\setval{S41}{0.173}%
\setval{M1}{0.4}%
\setval{M2}{0.7}%
\setval{Y1}{49}%
\setval{Y2}{0.501}%
\setval{Y3}{0.499}%
\setval{Y4}{+0.002}%
\setval{Y5}{\textminus{}0.002}%
\setval{Y6}{0.006}%

\newcommand{\papertitle}{In-context learning predicts clinical endpoints from bulk transcriptomes across diseases and
platforms}

\hypersetup{pdftitle={\papertitle},pdfauthor={Han Wang; Lennart Purucker; Huafeng Liu; Ziyin Luo; Junfeng Hao; Chao Zhu; Chunjie Tian}}

\usepackage{longtable,booktabs,array}
\begin{document}
\begin{center}
{\Large\bfseries \papertitle\par}
\bigskip
{\large Han~Wang\textsuperscript{1,*}, Lennart~Purucker\textsuperscript{2},
Huafeng~Liu\textsuperscript{3}, Ziyin~Luo\textsuperscript{4,5},
Junfeng~Hao\textsuperscript{6,7}, Chao~Zhu\textsuperscript{8,9},
Chunjie~Tian\textsuperscript{1,6,10,*}\par}

\end{center}

\begingroup
\small
\setstretch{1.0}
\setlength{\parindent}{0pt}
\setlength{\parskip}{4pt}
\textsuperscript{1}Department of Otorhinolaryngology, Affiliated Hospital of Guangdong Medical University, Zhanjiang 524000, Guangdong, China.\par
\textsuperscript{2}Prior Labs, Freiburg im Breisgau 79098, Germany.\par
\textsuperscript{3}Guangdong Provincial Key Laboratory of Autophagy and Major Chronic Non-communicable Diseases, Zhanjiang 524001, Guangdong, China.\par
\textsuperscript{4}Department of Otorhinolaryngology Head and Neck Surgery, The Sixth Affiliated Hospital, Sun Yat-sen University, Guangzhou 510655, China.\par
\textsuperscript{5}Department of Medicine, University of Pennsylvania, Philadelphia, Pennsylvania, USA.\par
\textsuperscript{6}Guangdong Provincial Key Laboratory of Autophagy and Major Chronic Non-communicable Diseases, Key Laboratory of Prevention and Management of Chronic Kidney Disease of Zhanjiang City, Institute of Nephrology, Affiliated Hospital of Guangdong Medical University, Zhanjiang 524001, Guangdong, China.\par
\textsuperscript{7}Department of Family Medicine, Shengjing Hospital of China Medical University, Shenyang 110022, Liaoning, China.\par
\textsuperscript{8}Department of Rheumatology and Immunology, The Third Affiliated Hospital of Naval Medical University, Shanghai 201805, China.\par
\textsuperscript{9}Department of Tropical Diseases, Naval Medical University, Shanghai 200433, China.\par
\textsuperscript{10}Otolaryngology Research Center, Guangdong Medical University Affiliated Hospital, Zhanjiang 524000, Guangdong, China.\par
\medskip
\textbf{Author contact information}\par
Han Wang\textsuperscript{*}: \href{mailto:han@hanziwww.com}{\nolinkurl{han@hanziwww.com}}.\par
Lennart Purucker: \href{mailto:lennart@priorlabs.ai}{\nolinkurl{lennart@priorlabs.ai}}.\par
Huafeng Liu: \href{mailto:liuhf@gdmu.edu.cn}{\nolinkurl{liuhf@gdmu.edu.cn}}.\par
Ziyin Luo: \href{mailto:luozy29@mail2.sysu.edu.cn}{\nolinkurl{luozy29@mail2.sysu.edu.cn}}.\par
Junfeng Hao: \href{mailto:ygzhjf85@gmail.com}{\nolinkurl{ygzhjf85@gmail.com}}.\par
Chao Zhu: \href{mailto:zhuchaokidney@163.com}{\nolinkurl{zhuchaokidney@163.com}}.\par
Chunjie Tian\textsuperscript{*}: \href{mailto:tianchunjie@gdmu.edu.cn}{\nolinkurl{tianchunjie@gdmu.edu.cn}}.\par
\textsuperscript{*}Corresponding authors: Han Wang and Chunjie Tian.\par
\endgroup
\clearpage

\begin{abstract}
Bulk transcriptomes carry information on diagnosis, treatment response and survival, but clinical cohorts are small,
and each gets a model of its own that often fails in the next cohort. Here we present PRTS, a
\val{D25}-million-parameter model that reads a labelled cohort as context and predicts binary, multiclass, continuous
or survival endpoints for new samples from whole transcriptomes, without per-cohort training. PRTS was pretrained on
\val{D16} million tasks generated from public RNA-seq profiles. On \val{D6} held-out
tasks and \val{D7} tasks from The Cancer Genome Atlas (TCGA), it had the highest mean predictive accuracy, narrowly
ahead of TabPFN v3.5, the strongest pre-specified comparator. The lead came from within-cohort prediction and, on
TCGA, mainly from survival tasks; across cohorts and at unseen sites, PRTS was level with the strongest tabular
models while running more than twice as fast as TabPFN v3.5. The same weights transfer labels to new cohorts from
public references, forecast accuracy from pilots, flag suspect labels and attribute predictions to genes. We release
the prior, the pretraining code and the model with its weights, together with RealBulk, a collection of \val{D1}
curated public cohorts, and its benchmark.

\end{abstract}

\section*{Introduction}
Bulk transcriptomes of tissue and blood carry information on diagnosis, disease subtype, treatment response and
survival, and public repositories now hold the expression profiles of hundreds of thousands of
people\cite{barrett2013,lachmann2018,weinstein2013}. Yet the cohorts that record any one clinical endpoint are small:
in the collection we assembled, the median cohort has \val{D10} people. Each cohort and endpoint gets its own model,
fitted from thousands of genes and about a hundred samples, and such models often fail when moved to another cohort,
platform or laboratory\cite{simon2003,leek2010,bernau2014}. Even in breast cancer, where expression signatures reached
the clinic\cite{vantveer2002}, most random gene sets are also associated with outcome\cite{venet2011}. Every new
cohort starts over, and what thousands of other cohorts could teach goes unused.

Foundation models offer two partial answers. Transcriptomic foundation models such as Geneformer, scGPT and
scFoundation learn gene representations from millions of single-cell profiles, and BulkFormer does so for bulk
profiles\cite{theodoris2023,cui2024,hao2024,bulkformer}, but every new endpoint still needs a model trained on its own
labelled samples. Tabular foundation models learn in context instead\cite{brown2020}: TabPFN and its successors learn
from millions of generated tables to read a labelled table together with new samples and predict in one forward pass,
with no training per task\cite{muller2022,hollmann2023,hollmann2025,qu2025,tabiclv2,tabpfn35}. In-context prediction
suits small samples, but tabular foundation models were designed for tables with far fewer features than a
transcriptome, carry no knowledge of expression data and, apart from survival-specific
variants\cite{survivalpfn,sic2026}, do not predict censored outcomes. They also assume that the labelled table and the
new samples come from one distribution, whereas a new cohort differs from the old in patients, site and platform.
GeneICL has recently brought in-context prediction to bulk expression\cite{geneicl}, and we compare PRTS with it
throughout.

Here we present PRTS (Prior-trained RNA-seq Task Solver), a model that predicts clinical endpoints from bulk
transcriptomes in context (Fig.~1). PRTS was pretrained on prior-generated tasks built from \val{D14} public RNA-seq
profiles, each anonymized so that the model never sees gene names, study identifiers or label names. Because genes
enter only through random codes, one model reads whatever genes a cohort measures, on microarrays or by RNA-seq. It
predicts binary, multiclass, continuous and censored survival endpoints, each with a full predictive distribution.
We curated RealBulk, a collection of \val{D1} public cohorts with clinical and phenotypic endpoints.

We tested PRTS on \val{D6} tasks from \val{D4} held-out RealBulk cohorts and on \val{D7} tasks
from \val{D13} projects of The Cancer Genome Atlas (TCGA) never used in development, against a primary family of
\val{B51} comparators and, post hoc, further configurations of TabPFN v3.5. The benchmarks separate three
situations: new patients from a cohort and sites that the context represents, patients from sites absent from it,
and a new cohort. PRTS has the highest mean accuracy on both benchmarks, measured as the rank correlation between its
predictions and the recorded outcomes, and it runs more than twice as fast as TabPFN v3.5. Its lead is narrow and
comes from prediction within a cohort. When the new samples come from another cohort, or from TCGA sites absent from
the context, PRTS is level with TabPFN v3.5, the strongest tabular foundation model, and the shift between cohorts
remains unsolved. Learning in context also opens applications that usually need models of their own (Fig.~1e). Without retraining, the same weights transfer labels to an unlabelled cohort from public reference
cohorts, forecast from a pilot the accuracy a full study will reach, flag samples whose labels are probably wrong and
attribute predictions to genes. We release the prior, the pretraining code and the model with its weights, together
with RealBulk and its benchmark, as a shared test bed for models of this kind.

\section*{Results}
\subsection*{PRTS predicts clinical endpoints from bulk transcriptomes in context}
To PRTS, a prediction task is a table: labelled context samples and unlabelled query samples, each with the expression
of every measured gene (Fig.~1a). PRTS reads the whole table in one forward pass and returns a predictive
distribution for each query sample, with no per-task training (Methods). A model that predicts in this way has to
learn, before it meets any cohort, what prediction tasks in expression data look like. PRTS learns from prior-generated
tasks built from PreBULK, \val{D14} public RNA-seq profiles of \val{D15} genes\cite{lachmann2018}.
RealBulk provides \val{D1} curated cohorts with clinical or phenotypic endpoints for development and evaluation
(Supplementary Fig.~1; Methods).

Pretraining supplies a volume and variety of tasks that a thousand curated cohorts cannot. A prior
samples profiles and genes from PreBULK and makes a target in one of four ways (Supplementary Fig.~2): from recorded
sample annotations, a masked gene or gene-set score, a causal mechanism rooted in real genes, or synthetic programmes
mapped onto real gene distributions. It then re-expresses the values in one of 19 formats and anonymizes the task, so
that nothing the model learns can depend on the identity of a gene. PRTS was pretrained on \val{D16} million such
tasks in three passes of increasing width, the last over tables of \val{D30} or more features (Methods).

The network has to read tables whose genes change from cohort to cohort. It has \val{D25} million parameters and no
gene identities (Supplementary Fig.~3): each gene enters as a random code, drawn afresh for every forward pass, and
query samples attend only to the context, so they never influence one another. Because attention along genes costs
time linear in their number, PRTS reads whole transcriptomes. At prediction time it averages four such passes, each
with its own gene codes and, for classification, its own order of the classes (Methods).

A fair test needs cohorts that the model has never seen. RealBulk's cohorts hold \val{D8} people in \val{D5} disease
groups, each cohort curated by a script, with verbatim labels and one sample per person (Fig.~1b). We set aside
\val{D4} of them, with \val{D9} people, as held-out cohorts, keeping cohorts linked by shared
people, accessions or publications on one side of the split. Public metadata do not always declare shared patients,
so we audited the split with further evidence (Methods and Supplementary Note~2). The linkage audit placed \val{K1}
held-out cohorts in a link component with a development cohort (Supplementary Tables~1 and 2), and we report the main
RealBulk comparisons both on all \val{D6} benchmark tasks and on the \val{C1} tasks that touch none of them.

Two benchmarks follow. The held-out cohorts define the RealBulk benchmark: \val{D6} tasks from microarray and RNA-seq
cohorts, \val{D11} within one cohort and \val{D12} across cohorts (Fig.~1c). The TCGA benchmark adds \val{D7} tasks
from \val{D13} projects never used in development, with endpoints from clinical records, survival and DNA; none is
derived from RNA. Together they cover three situations a user can be in: new patients from a cohort or site that
the context already represents, patients from a site absent from the context, and a new cohort. We score each task by
the rank correlation between predictions and recorded outcomes, which is zero for chance and one for a perfect
ranking. For a binary endpoint it equals 2$\cdot$AUC~$-$~1 and for a survival endpoint 2$\cdot$C~$-$~1, so that \val{M1}
corresponds to an area under the receiver operating characteristic curve (AUC), or a concordance index (C), of
\val{M2} (Methods). The primary family of comparisons pits PRTS against
\val{B51} comparators, from linear models to in-context models such as TabPFN v3.5 and GeneICL (Fig.~1d); all but
GeneICL were fixed before any result. Post hoc, we added further configurations of TabPFN v3.5, with more genes or
with two survival extensions, and Mitra, each configured on development tasks and analysed as a separate family
(Methods).

\subsection*{PRTS has the highest mean accuracy on both held-out benchmarks}
PRTS is the most accurate method on average on both benchmarks (Fig.~2a). Its margin over the strongest comparator,
TabPFN v3.5, is small. The margin holds on the tasks that share no people with development cohorts, narrows against
configurations of TabPFN v3.5 chosen post hoc, and on TCGA comes mainly from the survival tasks.

On the RealBulk benchmark, PRTS reached a mean rank correlation of \val{B1} and led TabPFN v3.5 by \val{B2}
(\ci{\val{B3}}{\val{B4}}; Holm-adjusted \Pv{} = \val{B5}; Fig.~2a,b). After Holm adjustment it led \val{B74} of the
\val{B51} comparators of the primary family; the exception, DeepSurv, which predicts only survival, was level with
PRTS on the survival tasks. On the TCGA benchmark, PRTS reached \val{B7} and led every comparator of the primary
family, TabPFN v3.5 by \val{B8} (\ci{\val{B28}}{\val{B29}}; Holm-adjusted exact sign-flip \Pv{} = \val{B9}).

Two checks bound this result. On the \val{C1} clean RealBulk tasks, PRTS led TabPFN v3.5 by \val{C3}
(\ci{\val{C4}}{\val{C5}}; Supplementary Fig.~4a), so the shared people found by the linkage audit do not carry it. On
TCGA, the lead rests mainly on the survival tasks: PRTS led TabPFN v3.5 by \val{T4} (\ci{\val{T5}}{\val{T6}}) on them
and was level with it on the other tasks (\val{T1}; \ci{\val{T2}}{\val{T3}}).

A stronger configuration of TabPFN v3.5 might close the gap, so we tested those a practitioner might choose. They
narrowed the margin on RealBulk without reversing it. PRTS led the \val{S37}-gene configuration by \val{S1}
(\ci{\val{S2}}{\val{S3}}) and the configuration chosen on development tasks, \val{B58} genes with SurvivalTabPFN (a
survival extension of TabPFN) for survival, by \val{S6} (\ci{\val{S7}}{\val{S8}}). Neither lead survives Holm
adjustment over the \val{S36} post hoc comparisons (\Pv{} = \val{S4} and \val{S9}; Supplementary Fig.~4c and
Supplementary Table~3), and PRTS was level with the \val{S37}-gene configuration given a discrete-time survival
extension (\val{S10}; \ci{\val{S11}}{\val{S12}}). On TCGA, PRTS led each configuration of TabPFN v3.5 run there.

The lead is spread over task types, with one exception. PRTS had the highest mean in every type on TCGA and in
three of the four on RealBulk (Fig.~2c and Supplementary Fig.~5a,b), although \val{B93} of the 8 type-level intervals
against TabPFN v3.5 include zero. The exception is survival on RealBulk, where a random survival forest reached
\val{B46} against \val{B45} for PRTS and SurvivalTabPFN was level with PRTS (\val{S14}; \ci{\val{S15}}{\val{S16}}).
On TCGA survival tasks, PRTS led SurvivalTabPFN by \val{S18} (\ci{\val{S19}}{\val{S20}}).

Means can hide what happens task by task, so we also counted wins. No method wins every task, but PRTS ranks first
most often: on \val{B25} of the RealBulk tasks and \val{B26} of the TCGA tasks, against \val{B38} and \val{B39} for
the runner-up (Supplementary Fig.~6). It is also first in every view of Elo-scale Bradley--Terry ratings from per-task
games\cite{bradley1952} (Fig.~2d and Supplementary Fig.~7a). The pre-specified analysis ran PRTS without reordering
the classes, a step adopted as the default after the benchmark. It reached the same conclusions; there, the
difference from TabPFN v3.5 on RealBulk was \val{B27} (Supplementary Fig.~7b; Methods).

\subsection*{PRTS leads within cohorts and is level with the strongest models elsewhere}
The lead is not uniform across the three situations. PRTS leads when the query patients come from a cohort and site
that the context already represents; when they come from a new cohort or a new site, it is level with the best
comparators. At every context size it had the highest mean, its survival curves scored best among those we
measured, and it ran at less than half the cost of TabPFN v3.5 (Fig.~3).

Within cohorts, on the \val{D11} tasks of that kind, PRTS led TabPFN v3.5 by \val{B32} (\ci{\val{B83}}{\val{B84}};
Fig.~3c; this breakdown by setting was added after the protocol was fixed). Across cohorts, with a source cohort as
context and a target cohort as query, it reached \val{B11} without a single target label and was level with TabPFN
v3.5 (\val{B75}; \ci{\val{B76}}{\val{B77}}) and with L2-penalized regression (\val{B78}; \ci{\val{B79}}{\val{B80}}).
A few labelled target samples helped PRTS and TabPFN v3.5 alike. With 5, 10 and 20 of them in the context, the rank
correlation of PRTS rose steadily to \val{B14}, while its difference from TabPFN v3.5 stayed between
\val{B81} and \val{B82}, every interval including zero (Fig.~3b and Supplementary Fig.~8b). These cross-cohort
comparisons depend on how a comparator chooses its genes: we rank them before each cohort is standardized, because
afterwards every gene has unit variance and the top \val{B58} are effectively arbitrary (Methods; Supplementary
Note~1 and Supplementary Table~4 give the results under both choices).

Sites behave like cohorts. In TCGA, a predictor that knows only each patient's tissue-source site reached a median
rank correlation of \val{G1} on the primary projects, so random folds partly test known sites. On the \val{G3} tasks
whose folds could be grouped by site, PRTS led TabPFN v3.5 by \val{G5} (\ci{\val{G6}}{\val{G7}}) with random folds.
When the query patients came from sites absent from the context, the margin fell to \val{G8}
(\ci{\val{G9}}{\val{G10}}; Fig.~3c and Supplementary Fig.~4b). Every method's mean fell on new sites, although for
most methods the interval of the change includes zero; that of PRTS changed by \val{G15} (\ci{\val{G16}}{\val{G17}})
and that of TabPFN v3.5 by \val{G18} (Supplementary Table~5). PRTS remained ahead of the linear, tree-based and
survival-specific methods.

Small cohorts are the common case, so we asked how accuracy depends on the size of the context. In the \val{B54} TCGA
tasks large enough for every size (\val{B68} of them from projects once used to compare checkpoints), we subsampled
the context to \val{B55}, \val{B56} or \val{B57} people. PRTS had the highest mean rank correlation at every size
(Fig.~3a and Supplementary Fig.~8a), from \val{B15} at \val{B55} people, against \val{B16} for TabPFN v3.5, to
\val{B36} with the full context, against \val{B37}. The margin over TabPFN v3.5, which also learns from a prior, was
no larger at \val{B55} people than with the full context, and these curves are descriptive.

The same order holds across disease areas and benchmark strata, with two caveats. The mean of PRTS was above those of
TabPFN v3.5 and GeneICL in cancer and in other diseases and on TCGA projects never or once used in development
(Supplementary Fig.~9a,b). The intervals against TabPFN v3.5 include zero in other diseases, and those against both
on the tasks whose target concept no development cohort carries.

A survival curve or a class probability says more than a ranking does, so we also scored the predicted distributions.
PRTS had the best integrated Brier skill score for survival on both benchmarks, \val{B23} on RealBulk and \val{B41}
on TCGA, against \val{B88} and \val{B89} for the best comparator, in both cases a post hoc configuration of TabPFN
v3.5 (Fig.~3d); these comparisons are descriptive. Of the 7 other probabilistic scores, PRTS was best on \val{B90} on
RealBulk and \val{B91} on TCGA, and TabPFN v3.5 on the rest (Supplementary Fig.~9c).

Accuracy has to be weighed against cost. PRTS took \val{B19} s per 1,000 samples on RealBulk against \val{B20} s for
TabPFN v3.5, which makes it \val{B21} times faster, and it was \val{B22} times faster on TCGA (Fig.~3e and Supplementary
Fig.~8c). Reordering the classes adds no time, because the four forward passes are run either way (Supplementary
Fig.~7c).

\subsection*{The Hub transfers labels to a new cohort from public reference cohorts}
The first application is label transfer. A new cohort often lacks the label a study needs, while public cohorts of the
same disease carry it. In a confirmation round on held-out targets, several such cohorts read together transferred
labels better than a single one, and PRTS did so as well as the strongest comparators given the same references; the
entropy of its predictions tracks how well a transfer works (Fig.~4).

We call this application the Hub, after the hub of labelled reference cohorts that it keeps for each endpoint
(Fig.~4a). On the benchmark, PRTS transferred from one source cohort as well as the best comparators; the Hub reads
several at once. For one endpoint, it draws its references from RealBulk development cohorts, reads eight random draws of two to four of them as context, with the unlabelled target cohort as
query, and averages the predictions (Methods). The settings of every application were fixed on development cohorts before its held-out
runs; the Hub and Explain were evaluated in two rounds, revised on development data in between, the second a
confirmation on held-out items that no earlier run had touched (Methods).

Several references beat one, but given the same references the strongest comparators did as well as PRTS. In the
confirmation round, on
\val{H18} held-out targets in \val{H19} task families, the Hub was ahead of PRTS reading a single reference cohort by
\val{H11} (\ci{\val{H12}}{\val{H13}}; Fig.~4b). It was narrowly ahead of GeneICL given the same references (\val{H3};
\ci{\val{H9}}{\val{H10}}; Fig.~4e gives the task types). It was level with TabPFN v3.5 given the same references
(\val{H2}; \ci{\val{H7}}{\val{H8}}) and with an L2-penalized model fitted on the pooled references, its penalty tuned
by inner cross-validation (\val{H14}; \ci{\val{H15}}{\val{H16}}). On the \val{H25} targets with classes, the Hub was
ahead of a SingleR-style correlation transfer\cite{aran2019} by \val{H1} (\ci{\val{H5}}{\val{H6}}), as it had been in
the first round. On targets whose cohorts share no people with development cohorts, its margins over TabPFN v3.5 and
pooled L2 were larger than on all targets in the confirmation and smaller in the first round (Supplementary
Table~6), so shared people do not consistently favour the Hub.

A user also needs to know when a transfer can be trusted. The entropy of the Hub's predictions is a guide: across
targets with classes, the lower the entropy, the higher the accuracy the Hub went on to reach (Spearman's
$\rho$~=~\val{H4} in the confirmation and \val{H17} in the first round; Fig.~4c). Published signatures set a
different bar, because each was derived for the endpoint it scores. On four held-out cohorts scored with three
published signatures\cite{cuzick2011,ifn2017,sweeney2016}, the Hub, reading every gene with nothing but the
references, scored above the locked published score in all \val{H23}, although only one of the four intervals
excludes zero (Fig.~4d).

\subsection*{The Planner forecasts, from a pilot, the accuracy of a full study}
The second application is study planning: guessing, usually from a small pilot, the accuracy a full study will reach.
From pilots of \val{P16} people, the forecasts of PRTS were closer to the outcome than an extrapolation of the pilot's
learning curve alone, and a permutation test checked beforehand whether the endpoint could be predicted at all
(Fig.~5).

We call this application the Planner. It computes the pilot's learning curve in context, letting 10 to \val{P18}
of its people form the context, and extrapolates the curve to the full size with an inverse power
law\cite{mukherjee2003,figueroa2012} (Fig.~5a). The law's prior for each task type was learned on development cohorts,
and the 80\% intervals come from the errors the Planner made there.

We then compared the forecasts with what the full cohorts gave. On \val{P10} held-out tasks, each with
five pilots of \val{P16} people, the Planner forecast the rank correlation PRTS reaches at the full size with a mean
absolute error of \val{P1}, against \val{P2} for least-squares extrapolation\cite{mukherjee2003} (Fig.~5b,c). The
difference is \val{P3} (\ci{\val{P5}}{\val{P6}}), an error ratio of \val{P7}. Its error was also lower than that
of a single prior pooled over task types (\val{P8}) and of the pilot's own rank correlation taken as the forecast
(\val{P9}). It was below that of least squares in every task type and in every disease area, some of which hold few
tasks (Fig.~5d), and its 80\% intervals covered \val{P4} of realized outcomes (Fig.~5b).

A forecast presumes that the label can be predicted at all, so the Planner tests that first, comparing the
out-of-fold rank correlation with that under permuted labels (Fig.~5e). At \Pv{} $\le$ \val{P17}, it rejected
\val{P12} of \val{P13} tasks with shuffled labels, close to the nominal rate, and \val{P14} of \val{P15} tasks with
real labels.

\subsection*{Audit flags suspect labels and Explain finds the genes behind predictions}
The last two applications look inside a cohort and inside a prediction. Audit ranked samples with wrong labels more
precisely than a random forest did, and the attributions of Explain identified the genes that PRTS relies on and,
where DNA defines the truth, pointed to it (Fig.~6).

The third application is a label audit. Labels in public cohorts are sometimes wrong, and a wrong label hides among
hundreds of correct ones\cite{frenay2014}. Audit uses the out-of-fold predictions of PRTS to rank a cohort's samples
by how likely their recorded label is wrong (Fig.~6a). We flipped \val{A13} of labels to another class at random in
\val{A9} held-out classification tasks; Audit ranked the flipped samples with an average precision of \val{A1},
against \val{A2} for a random forest and \val{A14} for L2-penalized logistic regression (Fig.~6b). Its lead over the
random forest was \val{A3} (\ci{\val{A4}}{\val{A5}}) with cohort clusters weighted equally and \val{A6}
(\ci{\val{A7}}{\val{A8}}) with disease areas weighted equally (Fig.~6c).

The fourth application, Explain, asks which genes a prediction rests on. It attributes each prediction to genes by
integrated gradients\cite{sundararajan2017} and scores gene sets from the attributions (Methods). The attributions
identify the genes PRTS relies on. On \val{E4} tasks with planted gene sets, masking the top 10 genes of PRTS changed
its predictions \val{E23} times as much (\ci{\val{E24}}{\val{E25}}) as masking the top 10 genes of moderated
$t$\cite{smyth2004}. On \val{E9} TCGA chromosome-arm tasks, the ratio was \val{E26} (\ci{\val{E27}}{\val{E28}};
Fig.~6d), and in the first round masking the top genes of PRTS changed its predictions far more than masking 10
random genes (Fig.~6e).

Where DNA defines the truth, the attributions point to it. In the chromosome-arm tasks, the label is a tumour's
arm-level copy-number call and the arm's genes are the truth. The true arm ranked near the top for both methods, at a
rank percentile of \val{E14} for PRTS and \val{E15} for moderated $t$ (difference \val{E11};
\ci{\val{E12}}{\val{E13}}; Fig.~6f). Outside cancer, we paired each of \val{E21} disease-module tasks in advance with
an established Hallmark programme\cite{liberzon2015}. PRTS ranked the expected programme at a mean rank of \val{E18}
among \val{E22} (Fig.~6g), against \val{E19} to \val{E20} for moderated $t$, preranked gene set enrichment analysis
(GSEA)\cite{subramanian2005} on moderated $t$, L2 coefficients and LightGBM Shapley additive explanation (SHAP)
values\cite{lundberg2017}.

\section*{Discussion}
One model, pretrained on prior-generated tasks, predicts clinical endpoints from bulk
transcriptomes at least as accurately as the strongest tabular foundation models, with no gene selection or per-task
tuning. On two held-out benchmarks that share no cohort and were curated separately, PRTS was the most accurate method
on average and ran more than twice as fast as TabPFN v3.5. Its lead over TabPFN v3.5 came from prediction within
cohorts, and on TCGA it rested mainly on the survival tasks. Across cohorts and on new TCGA sites, PRTS was level with
the best comparator rather than ahead of it, and on RealBulk its leads over TabPFN v3.5 configured on development
tasks or reading \val{S37} genes did not survive adjustment for the post hoc comparisons. The advantage of PRTS thus
lies in learning from a context that shares the query's cohort and sites, not in bridging between them.

We did not ablate the design, so we cannot apportion the lead between the prior, the gene codes, the distributional
heads. When no development cohort carried the target concept, PRTS was level with TabPFN v3.5 and GeneICL.
The design does explain why one model can serve so many
cohorts: random gene codes free PRTS from gene identity, and attention along genes through inducing tokens keeps the
cost linear in their number, so the same weights read a microarray or a whole transcriptome. Predicting full
distributions, censored event times included, yields survival curves, and these had the best integrated Brier skill
score we measured. Part of the within-cohort lead may rest on structure that patients of one cohort or site share:
when the query patients came from unseen sites, the mean rank correlation of PRTS changed by \val{G15}
(\ci{\val{G16}}{\val{G17}}), and its margin over TabPFN v3.5 shrank from \val{G5} to \val{G8}.

Because PRTS learns in context, the same weights support applications that usually need models of their own, and
their results follow the benchmark's pattern. For label transfer, reading several public cohorts at once was ahead of
single-reference and correlation-based transfer but level with TabPFN v3.5 and pooled L2-penalized regression given
the same references; the entropy of the predictions tracks how well a transfer works. For study planning, a pilot
becomes a forecast, with an interval, of the accuracy PRTS will reach on the full cohort, and a permutation test says whether
the endpoint is learnable at all. Samples can be ranked for label review before analysis, and predictions traced to
the genes the model relies on.

The evaluation has limits. Public cohorts can share patients without declaring it: a linkage audit with more evidence
than the original linkage tied \val{K1} held-out cohorts to development cohorts. The main comparisons held on the
\val{C1} tasks that touch none of them (Supplementary Fig.~4a), and one development endpoint proved mislabelled and is
now corrected in RealBulk. Random folds in TCGA mix new patients with known sites, and no method received a cohort or
site index, which models built for grouped data might use to advantage. We did not run every tabular model available,
hosted ones among them, nor hybrids such as a tabular foundation model reading BulkFormer's embeddings, which only L2
models read here. Newer tabular foundation models are likely to close the remaining gap. The cohorts are
retrospective.

RealBulk and its benchmark may outlast any one model. We release the curated cohorts, the task definitions with
their clean subset and site-grouped folds, every method's predictions and the evaluation code, so that new models can
be scored on the same tasks; we would welcome the tasks' inclusion in community benchmarks such as
TabArena\cite{tabarena}. We also release the prior that generates the pretraining tasks, the pretraining code and
the model with its weights, so that others can build on each part. Ablations of the prior, the gene codes and the
distributional heads come next, with larger and more diverse corpora, other molecular layers and prospective
clinical studies. The recipe (a prior built on real profiles, anonymized tasks and prediction in context) may also
carry over to other measurements with many features and few labelled samples.

\section*{Methods}

\subsection*{Tasks, endpoints and the accuracy measure}
A task pairs labelled context samples with unlabelled query samples for one endpoint. PRTS reads the expression
profiles and labels of the context and the expression profiles of the query, and returns a predictive distribution
for every query sample in one forward pass (four such passes are averaged; see Inference). Four task types are
supported: binary and multiclass endpoints (class
probabilities for up to 64 classes), continuous endpoints (a predictive distribution of the value) and right-censored
survival endpoints (a distribution of the event time).

Accuracy is measured as the rank correlation between predictions and recorded outcomes, which puts the four types
on one scale on which zero is chance and one is a perfect ranking. For binary tasks it is Somers'
$D$\cite{somers1962}, 2$\cdot$AUC~$-$~1, and for multiclass tasks its mean over classes, 2$\cdot$macro-AUC~$-$~1 (one
class against the rest). For continuous tasks it is Spearman's $\rho$ between the predictive median and the label,
and for survival tasks 2$\cdot$C~$-$~1, where C is Harrell's concordance index\cite{harrell1982}. AUC is the area
under the receiver operating characteristic curve. PRTS's survival risk is minus the logarithm of its predicted
median event time. A benchmark's mean rank correlation is the mean of the four task types' means, so that every type
weighs the same. The released code and tables store this measure in the column \texttt{rank\_skill}.

Within one cohort, a task is a stratified five-fold partition of the cohort's people, seeded from the task
identifier. Each fold serves once as the query, with the other four as the context, and the pooled out-of-fold
predictions are scored. Across cohorts, the context is every labelled person of a source cohort and the query every
labelled person of a target cohort of the same task family; labels pass through the family's class map, and only
genes measured in both cohorts are used. In the partial-target design, 5, 10 or 20 labelled target samples join the
context (three repeats each) and the remaining target samples are scored. In the few-sample design, the context of
each TCGA fold is subsampled to 32, 64 or 128 people (three repeats per fold), and a size counts only where it is
smaller than the fold's context. The curves use the \val{B54} TCGA tasks, of all 64, in which every size is available:
\val{B67} from projects never used in development and \val{B68} from projects once used to compare checkpoints.
Repeats are scored one by one and then averaged.

\subsection*{Architecture}
PRTS is a transformer\cite{vaswani2017} that treats a task as a table of cells, one per sample and gene, and predicts
from the whole table in context\cite{muller2022,hollmann2025}. It has \val{D18} parameters (Supplementary Fig.~3).
Its design builds on TabPFN v2, which attends along samples and features and gives features random
codes\cite{hollmann2025}, and on TabICL, which embeds features through set attention with inducing points and
compresses each sample into a row before predicting in context\cite{qu2025}.
Within one cohort, PRTS reads every gene the cohort measures on the natural-log scale and standardizes each gene on
the context samples only, so no query value enters a statistic. Across cohorts, each cohort is first z-scored over its
own samples. Values are soft-clipped, and a missing value becomes zero with a missingness indicator.

\textbf{Cell tokens.} Each cell becomes a 192-dimensional token, the sum of four parts. The first encodes the
standardized value (the value and 16 radial basis functions with learned centres and widths). The second is a random
code of the gene, a 32-dimensional Gaussian vector scaled by 0.1 and projected; it is drawn afresh for every forward pass,
so that no gene identity can be learned\cite{hollmann2025}. The third embeds the label on context samples, or a learned
\emph{unknown} token on query samples\cite{tabiclv2}; the fourth embeds the task type. Each class has a learned
vector; a continuous label enters as its value standardized on the context, and a survival label as its standardized
log time, its quantile within the context and the event indicator.

\textbf{Cell blocks.} Four cell blocks follow. Each applies attention along samples within a gene, a feed-forward
layer (SwiGLU\cite{shazeer2020}, expansion two) and attention along genes within a sample, with six attention heads
and every sub-layer with RMS normalization\cite{zhang2019} and a residual sum. In attention along samples, every sample attends only to the
context samples and to eight learned gene summaries, so query samples never influence the context or one another.
Attention along genes passes through 128 inducing tokens\cite{lee2019} and four routing tokens, so its cost grows
linearly with the number of genes.

\textbf{Row blocks and heads.} Eight learned summaries of 128 dimensions read each sample's cells by cross-attention and
are concatenated into one 1,024-dimensional row. Twenty row blocks (16 heads) then let query rows read context rows;
these blocks hold 96\% of the parameters. A readout feeds three heads: class probabilities from a hybrid of a
multilayer perceptron and retrieval over context rows; 999 quantiles of a continuous target, standardized on the
context, with exponential tails beyond the outermost quantiles; and a mixture of eight log-normal distributions of the
event time for survival.

\subsection*{The prior: tasks built from public expression profiles}
\textbf{PreBULK.} The prior draws expression from PreBULK: \val{D14} human RNA-seq profiles of \val{D15} genes from
ARCHS4 (version 2.7)\cite{lachmann2018}, as ln(TPM~+~1). A profile enters PreBULK only from a transcriptomic library
with at most 70\% zero values, from a study of at least 16 profiles, and if it duplicates no other profile. From
PreBULK the prior fits its assets: 64-point quantile tables of every gene, 200 co-expression modules over 4,000 genes
and 128 principal components.

\textbf{Sampling the features.} A task's samples come from one study or several (16 to 512 samples; 15\% of tasks
are taller, up to 2,048). Its genes are a subset of 32 to 20,010 genes, with widths drawn equally often from four
ranges (32--256, 256--2,048, 2,048--8,192 and 8,192--20,010). The subset follows one of five rules: uniform over
genes; highly variable genes of the sampled profiles; whole co-expression modules; root genes hidden among decoys
matched for detection and variance; or a mixture of detection-rate strata that includes sparse genes. Tasks built
from RealBulk development cohorts sample a cohort's people and a subset of the genes it measures in the same way,
after harmonization onto PreBULK's gene universe and log scale (Supplementary Fig.~2a).

\textbf{Targets.} One of four branches makes the target. Annotation targets take recorded sample annotations from the
public metadata (2 to 20 classes, at most 90\% in the largest class, identifier-like fields excluded). Self-supervised
targets are a masked gene, a gene-set score or a principal-component score, with the target genes and the genes
correlated with them removed. Mechanistic targets come from random causal graphs of 2 to 32 nodes rooted in real
genes, modules, principal components or covariates, with 14 families of functions and 7 of noise; a solvability level
sets the share of explained variance and of uninformative genes. Fully synthetic tasks draw expression from latent
programmes and sparse mixtures and map it onto real gene marginals through the quantile tables. A head then gives the
target a task type: classes from score bins, a softmax, prototypes or tree regions; continuous values with optional
heavy tails, mixtures, heteroscedastic noise, counts or bounds; and event times from accelerated failure-time
(log-normal or Weibull) or Cox\cite{cox1972} models with 10--70\% censoring.

\textbf{Re-expression.} Every value distribution is re-expressed with parameters drawn per task. A task takes one
of 19 value formats, for example ln TPM, log2 FPKM, counts at a sequencing depth of $2\times10^{5}$ to $5\times10^{7}$
with negative-binomial dispersion of 0.05 to 1, microarray-like log2 intensities, z-scores, ranks or quantile
normalization; formats are mixed between studies in 15\% of tasks. Depth scaling, dropout, batch shifts, monotone
warps and quantization follow, with missing values (up to 20\%, completely at random, at random or not at random),
outliers, uninformative and duplicated genes, and covariates. A quarter of tasks add a stress design: a confounder, a shortcut,
a covariate shift or a grouped split.

\textbf{Anonymization.} Every task is anonymized before the model sees it. The gene order is shuffled, samples and
class labels are re-ordered, no gene names are kept, and each gene carries only a random code that changes from one
forward pass to the next. PRTS therefore sees no gene names, study identifiers or label names.

\textbf{Admission.} Four gates admit a task. Validity requires finite values, at least eight context samples and one
query sample, and a usable class structure. Learnability requires an extra-trees model whose out-of-bag predictions
beat the constant predictor in at least 95\% of 200 bootstrap resamples\cite{tabiclv2}. Discoverability applies to
wide tasks (at least 256 genes): in a 64-sample context, a root gene's marginal effect must exceed the 90th percentile
of matched decoys in two of three draws. A difficulty quota matches the accepted tasks' out-of-bag accuracy to a target
histogram of eight bins. Continuous tasks whose query values lay far outside the context's range were also excluded.

\subsection*{Pretraining}
PRTS was pretrained on \val{D23} prior-generated tasks in three passes of increasing width (Supplementary Fig.~2c).
Each pass took tasks within its limits from libraries generated as above. The first two passes held 2,559,793 tasks
with up to 2,048 genes and 2,560,000 tasks with up to 8,192 genes. The third, in which each task was used once, held
921,272 tasks of 10,000 to 22,011 features (genes plus duplicated genes and covariates) and 50 to 640 samples, drawn
in equal numbers from four sample-size bins.
The loss is the cross-entropy for classes; for continuous targets, the continuous ranked probability score, computed
as the mean pinball loss over the 999 quantiles; and for survival, the negative log-likelihood of the log-normal
mixture at the event times the prior draws before censoring. Two-dimensional backbone matrices were optimized with
Muon\cite{liu2025muon} (momentum 0.95, Nesterov updates, five Newton--Schulz iterations) and all other parameters with
AdamW\cite{loshchilov2019} ($\beta$ of 0.9 and 0.95, weight decay 0.01), with gradients clipped at a norm of 10,
bfloat16 arithmetic and 128 tasks per update on four NVIDIA RTX PRO 6000 GPUs. Each pass warmed up linearly (over 2,000, 500 and 500
updates) to a learning rate of $3\times10^{-4}$, $2\times10^{-4}$ and $5\times10^{-5}$, respectively, and then
decayed along a cosine to a tenth of it.

\subsection*{Inference}
At prediction time, PRTS averages four forward passes, each with its own random gene codes and, for classification,
its own random order of the classes; class probabilities and quantiles are averaged, and survival mixtures pooled.
The pre-specified analysis ran the four passes without reordering the classes; for continuous and survival tasks the
two variants coincide. The released code names the default, with class reordering, its \texttt{thinking} preset,
and the released tables label the variant without reordering `PRTS (Thinking off)'. Benchmark tables were capped at 24 million cells, a cap no
benchmark task reached.

\subsection*{RealBulk}
RealBulk collects \val{D1} public cohorts of human bulk RNA-seq and expression-microarray profiles with clinical or
phenotypic endpoints (\val{D8} people in \val{D5} disease groups; Supplementary Fig.~1), from the Gene Expression Omnibus (GEO)\cite{barrett2013} and
other open-access repositories. TCGA, and any collection that contains TCGA samples, is excluded. Every cohort was
curated by a script that records its provenance. Labels map the deposited values one to one, with no merged classes,
no imputation and no thresholds unless the source defines them. Endpoints derived from expression (molecular
subtypes, signatures, deconvolution) and demographic or technical variables are never targets. One sample is kept per
person: paired normal tissue, later visits and replicates are excluded. An endpoint needs at least 10 people per
class, at least 50 labelled people (30 in a relaxed tier of \val{D26} development cohorts, which also admits sorted
cell populations) and, for survival, at least 10 events. Releases of the same patients (SuperSeries, reanalyses) were
merged into one cohort, no sample accession occurs in two cohorts, and a scan of per-sample gene ranks across cohorts
resolved shared arrays and libraries. Expression was harmonized onto PreBULK's 20,010-gene universe on the natural-log
scale; for a gene measured by several probes, the probe with the highest mean was kept. Cohorts that measure the same endpoint
concept form a task family with a class map.

\subsection*{Held-out cohorts and leakage controls}
Cohorts linked by shared people, accessions, publications or a recorded overlap formed link components, which were not
split between development and evaluation\cite{kapoor2023}. Before evaluation, \val{D4} cohorts (\val{D9} people)
were set aside as held-out cohorts. All are strict-tier cohorts with one sample per person and at least 5,000 universe
genes. By the records available then, they had no sample accession in PreBULK and no recorded link to any cohort
used during development,
including the cohorts used to calibrate the prior and to compare pretraining checkpoints. The held-out data were stored apart from the
development data and never used to develop PRTS. The other \val{D2} usable cohorts (\val{D3} people) are the
development cohorts; three cohorts were unusable. Of the held-out cohorts, \val{D27} were profiled on microarrays and
\val{D28} by RNA-seq.

\textbf{Audit of the split.} To test the split independently of the original linkage, we rebuilt the link graph from
every record the curation kept: repeat rules, sample exclusion reasons, shared-array files, curation notes and survival times matched up
to a unit factor. We also used mutual-best expression matches between cohorts, under a rule calibrated so that at most
one in 40,000 pairs of cohorts from different disease groups reached it (Supplementary Note~2). The rebuilt graph links
\val{K8} cohort pairs, against \val{K7} in the original split. It places \val{K1} held-out cohorts in a component with
a development cohort, \val{K2} of them on strong evidence (a record naming shared people, a curation note or
expression matches; Supplementary Note~2) and \val{K3} on a recorded possibility of shared people. GSE17536, for example, is almost wholly contained in the
development cohort GSE14333. Removing the \val{K4} benchmark tasks that touch these cohorts, or whose source and target
cohorts share people, leaves the clean set of \val{C1} tasks; no task was added or reselected (Supplementary Tables~1 and 2).
Matching through repaired sample identifiers found four held-out RNA-seq cohorts with samples of their own in
PreBULK, none of them with a benchmark task; \val{K10} tasks include people with another sample in PreBULK (\val{K11} of them in the clean
set). Within \val{K14} held-out cohorts, \val{K13} pairs
of near-identical samples carry different person identifiers; they were left in place, so every method saw them.

Held-out tasks met quality floors. Binary tasks needed at least 15 people in the smaller class and 60 in all, and
multiclass tasks at least 12 per class and 80 in all. Continuous tasks needed at least 60 people and 10 distinct
values, and survival tasks at least 30 events in a cohort, or 20 in each cohort of a pair. Before any benchmark
result, a seeded rule selected \val{D6} of the 200 tasks defined on held-out cohorts (\val{D11} within one cohort and
\val{D12} across cohorts), keeping every cross-cohort task and balancing cancer and other diseases where both had
tasks. A target concept counts as new when no development cohort carries it.

\subsection*{TCGA benchmark}
The TCGA benchmark uses open-access data from the Genomic Data Commons\cite{weinstein2013,grossman2016}: STAR
gene-level TPM mapped onto the gene universe as ln(TPM~+~1), primary tumours only and one aliquot per patient, with
formalin-fixed samples, samples flagged against use, redacted cases and patients given neoadjuvant treatment excluded.
Endpoints come from clinical records, curated survival\cite{liu2018cdr} and DNA: tumour mutation burden from consensus
mutation calls\cite{ellrott2018}, and arm-level copy number and aneuploidy\cite{taylor2018}. No endpoint was derived
from RNA, and age and sex were not used. Each project contributed at most one task per type, the first entry of a fixed
priority list that met the RealBulk floors; the lists favoured overall survival, nodal or mutation status, stage or
grade, and log tumour mutation burden. TCGA tasks from 17 cBioPortal studies had
been used to compare pretraining checkpoints; the benchmark's primary stratum holds the \val{D13} projects never used
in this way (LUAD, LUSC, HNSC, LIHC, PRAD, CESC, ESCA, PAAD, GBM, KICH, MESO, THYM and UVM), with \val{D7} tasks. No RealBulk cohort contains TCGA samples. A content scan flagged any patient whose closest RealBulk or
PreBULK profile (Spearman correlation of the two profiles over the 5,000 most variable genes) was nearer than any two different patients of
its project; it removed \val{D20} patients, all for matches in PreBULK.

The benchmark's folds were drawn without regard to the tissue-source site, so most query patients share a site with
context patients. A site-only predictor, which gives each query patient the mean context label of its site
(survival through martingale residuals), measures how much the site alone reveals. Post hoc, we also drew
site-grouped folds wherever a task had at least eight sites and no site above 30\% of its patients. Their five folds
keep every site whole, so every query patient comes from a site absent from the context; sites were assigned largest
first so as to spread classes, events or target quintiles evenly. Such folds were feasible for \val{G2} of the
\val{G4} TCGA tasks, \val{G3} of them in the primary stratum. PRTS and the comparators
that predict every task type (TabPFN v3.5, TabICLv2, L2, elastic net, random forest and gradient boosting) and the
survival-specific methods ran on these folds with their benchmark settings.

\subsection*{Comparators}
Thirteen comparators ran on the same evaluation units (a fold, a cohort pair or a repeat), the same query people and
the same harmonization. Unless stated otherwise, comparators read the 2,000 genes of highest context variance,
z-scored on the context. Across cohorts, the genes are ranked by their variance over the context before each cohort is
standardized on its own samples, because that standardization leaves every gene with unit variance
(Supplementary Note~1). Classical
methods chose their settings by inner
cross-validation on the context (five stratified folds, three below 60 samples, grouped by person). They were L2-penalized logistic, ridge or
Cox regression; elastic net or Coxnet\cite{zou2005,simon2011}; random forests or random survival
forests\cite{breiman2001,ishwaran2008}; and gradient boosting (LightGBM\cite{ke2017}; XGBoost\cite{chen2016} for
survival)\cite{pedregosa2011,polsterl2020}. The deep-learning comparators were RealMLP-TD\cite{holzmuller2024} and
TabM-D\cite{gorishniy2024} with tuned defaults and, for survival, DeepSurv\cite{katzman2018,kvamme2019}. The in-context
comparators were TabPFN v3.5\cite{tabpfn35}, TabICLv2\cite{tabiclv2} and GeneICL\cite{geneicl}, with the
survival-specific SurvivalPFN\cite{survivalpfn} and SIC\cite{sic2026}. BulkFormer-147M\cite{bulkformer} supplied
label-free sample embeddings, read by the L2 models. TabPFN v3.5 (its multitask checkpoint, with the multiclass
checkpoint for three or more classes) and TabICLv2 (its classifier and regressor) read either the 2,000 most variable
genes or the 500 genes most associated with the context's labels. The choice went to whichever scored higher on 215
tuning tasks from development cohorts, before any benchmark run. A post hoc check on the \val{Y1}
multiclass tuning tasks found TabPFN v3.5's two checkpoints level: a mean rank correlation of \val{Y2} with the
multiclass checkpoint and \val{Y3} with the multitask one (difference \val{Y4}; \ci{\val{Y5}}{\val{Y6}}). Comparators without a survival output regressed the
martingale residuals of the context's Kaplan--Meier model\cite{kaplan1958}. For the probabilistic scores, methods
that give only a risk received survival curves from a one-dimensional Cox model with a Breslow baseline fitted on
cross-fitted context risks, and point predictions received quantiles from cross-fitted context residuals. Survival
methods that give only curves ranked patients by minus their restricted mean survival time. GeneICL ran in its authors' default configuration on every gene. Two further methods, $k$-nearest
neighbours and DeepHit\cite{lee2018deephit}, ran outside the pre-specified family on the 2,000 most variable genes. For the ratings, L2 was also run on every
measured gene, solved exactly in the row space of the context, and each leaderboard keeps the better of its two
configurations. Each method had 30 minutes per unit, and a failure scored as the null model (the context's class
shares, quantiles or Kaplan--Meier curve).

\textbf{Post hoc comparators.} To test whether TabPFN v3.5 was held back by its input or by its survival adapter,
we also ran the configurations a practitioner might choose, each set on the 215 development tuning tasks before its
held-out run. TabPFN v3.5 read 6,000 genes, or every gene, with enough
ensemble members of at most 768 genes each for every gene to be read. Two survival extensions of TabPFN v3.5 replaced
the martingale adapter: SurvivalTabPFN from tabpfn-extensions\cite{tabpfnext} and a discrete-time classifier on
person-period data\cite{kvamme2021}. Mitra\cite{mitra2025} read the 256 genes most associated with the context's
labels. On the tuning tasks, TabPFN v3.5 reached a mean rank correlation of \val{S30} with 2,000 genes,
\val{S31} with 6,000 and \val{S32} with every gene; on the 30 survival tuning tasks, SurvivalTabPFN reached \val{S33},
the martingale adapter \val{S34} and the discrete-time classifier \val{S35}. The configuration chosen on development
data is therefore 2,000 genes with SurvivalTabPFN for survival tasks. We also report a second
combination that was suggested to us, 6,000 genes with the discrete-time extension, although the tuning tasks favoured
neither part; both combined
configurations were assembled from the runs above, without new predictions. Mitra ran on
classification tasks only, because its regression head needs a training step on each task, which the in-context
protocol excludes.
The discrete-time extension's TCGA run was stopped, unread, when the set of comparators was frozen, and is not reported.
These comparators enter no pre-specified analysis; they are not rated, and their \Pv{} values are Holm-adjusted among
themselves.

\subsection*{Statistics}
The two benchmarks were never pooled for inference. For each comparator, the difference in mean rank correlation (PRTS
minus comparator) has a 95\% percentile interval from 10,000 bootstrap resamples\cite{efron1979} of clusters: the
\val{B69} link components of the RealBulk benchmark, a cross-cohort task forming a cluster with the pair of components
it joins, or the 13 TCGA projects. On the clean set, the clusters are the components of the rebuilt link graph. The
overall contrasts
carry \Pv{} values adjusted by Holm's method\cite{holm1979} over the 13 comparators of the primary family, GeneICL included, and separately
over the post hoc comparators. On RealBulk, the unadjusted \Pv{} is twice the smaller share of
resamples on either side of zero, with one added to numerator and denominator; on TCGA, it comes from an exact
sign-flip permutation test over all 8,192 sign patterns of the projects. The site-grouped analysis resamples and
sign-flips the nine primary projects that have feasible tasks. Intervals by task type and by stratum are
descriptive. A task counts as a win, tie or loss with a margin of 0.01. Elo ratings are Bradley--Terry\cite{bradley1952} maximum-likelihood ratings from per-task games between every
pair of methods that predicted the task: the higher rank correlation wins, a difference within 0.01 is a draw counted as
half a win, and 0.1 virtual draws join every pair. Ratings are on the Elo scale, centred at 1,500, with 95\% intervals
from 1,000 cluster bootstrap resamples; survival-only methods play only survival tasks. The all-tasks view pools both
panels, including the TCGA projects once used to compare checkpoints, and is descriptive. A method ranks first on a
task when no method has a higher rank correlation, so tied methods each count. Probabilistic scores were computed per task:
log loss (clipped at $10^{-15}$), the Brier score\cite{brier1950} and top-label expected calibration error over 10
bins\cite{guo2017}. For survival, the integrated Brier score used inverse-probability-of-censoring
weights\cite{graf1999} over 50 quantiles (10th to 90th percentile) of the context's follow-up times and is expressed
as a skill score: one minus its ratio to the score of the Kaplan--Meier null model, pooled over tasks by the geometric mean. A method's improvability is its mean gap to the best method's rank correlation
over tasks; the accuracy--cost front joins the methods that no other method beats on both improvability and time.
All methods ran on the same computing node, and time was recorded per unit; for PRTS, whose four GPUs were shared
between jobs, per-unit times were apportioned from each GPU's total in proportion to the units' cells. Costs use the
times of the original runs, because the cross-cohort units repeated after the gene-ranking correction ran under a
different load.

\textbf{Pre-specification and changes.} The benchmark protocol (panels, comparators, scoring and statistics) was fixed
before any result. The pre-specified analysis pairs every comparator with PRTS without class reordering; we report
PRTS with class reordering, adopted as the default after the benchmark, and show the pre-specified analysis in
Supplementary Fig.~7b. Outside the protocol are GeneICL, the Elo analysis, the contrasts by setting, the post hoc
comparators, the clean set and the site-grouped folds. Each post hoc comparator was configured on development tuning
tasks before its held-out run; the clean set was defined by the audit of the split, without reference to any result;
and the site-grouped folds follow a feasibility rule fixed before any method ran on them. The comparators' gene
ranking across cohorts was corrected after the first run of the benchmark and every affected run repeated; as the protocol requires
after a correction, Supplementary Note~1 and Supplementary Table~4 give both values of every contrast it changed.

\subsection*{Hub}
The Hub labels a cohort that has no labels of its own. For each endpoint concept, it keeps a set of labelled
reference cohorts, drawn from the development cohorts of a task family. For a target cohort, PRTS reads eight random draws of two
to four reference cohorts (at most 700 context samples per draw), each cohort z-scored on its own, as context and the
target cohort as query, and averages the eight predictions. The entropy flag is the mean normalized entropy of the
class probabilities over the target's samples. TabPFN v3.5 and GeneICL read the Hub's own reference draws, TabPFN
v3.5 on the 2,000 genes of highest variance within the reference cohorts before each was standardized
(Supplementary Note~1). The other
comparators were PRTS reading a single reference cohort and a SingleR-style transfer\cite{aran2019} (our
reimplementation: marker genes from per-class $t$ statistics, then a softmax over each target sample's Spearman
correlation with the class medians). A further comparator was an L2-penalized model fitted on the pooled references,
with its penalty tuned by inner cross-validation: five folds for classes, and efficient leave-one-out for continuous
targets and for survival targets, whose martingale residuals it regressed. Contrasts weight task families equally,
with 95\% intervals from family bootstrap resamples. On four held-out cohorts, three published signatures were scored by their locked published scores, computed from
the published genes and weights: the cell-cycle progression score in prostate cancer\cite{cuzick2011}, an
interferon score in systemic lupus erythematosus\cite{ifn2017} and a three-gene tuberculosis
signature\cite{sweeney2016}. The Hub was compared with each score using patient bootstrap intervals. Of the \val{H18} confirmation targets,
\val{H40} lie in cohorts that the audit of the split linked to development cohorts; the contrasts on the other
\val{H33} use the same family bootstrap (Supplementary Table~6).

\subsection*{Planner}
The Planner forecasts, from a pilot, the rank correlation PRTS will reach at a target size. PRTS computes the pilot's
learning curve in context: 10, 15, 20 and 25 of its people form the context (stratified by class or event status;
five draws) and the rest the query. The Planner fits an inverse power law, $s(m) = a - b\,m^{-c}$, to the curve by
importance sampling from a prior over $(a, b, c)$ for the task type, learned from full learning curves of development
cohorts, with a Gaussian likelihood of the curve's means\cite{mukherjee2003,figueroa2012}. Here $s$ is the rank correlation and $m$ the
number of context people. The forecast is the posterior
median at the target size, and its 80\% interval adds the type's 80\% quantile of absolute errors on development
cohorts. The comparators are least-squares extrapolation of the same law\cite{mukherjee2003}, one prior
pooled over task types, and the pilot's own rank correlation. Each held-out task (at least 100 people, one task per cohort) has a
full learning curve at eight sizes up to 85\% of its people (five draws), whose largest size is the target, and five
pilots of 30 people. The feasibility test asks whether a label is predictable at all: it compares a cohort's
out-of-fold rank correlation (five folds) with that of 19 permutations of the labels between people, so that
$\Pv = (1 + \text{number of permutations at least as good})/20$. Its size was measured on \val{P13} tasks with
shuffled labels and its power on \val{P15} tasks with real labels.

\subsection*{Audit}
Audit ranks a cohort's samples by how likely their recorded label is wrong. PRTS predicts every sample out of fold (10
folds), and a sample's score is one minus the predicted probability of its recorded label\cite{northcutt2021}. To
evaluate Audit, we flipped 10\% of the labels in each held-out classification task to another class uniformly at
random (seeded per task)\cite{frenay2014} and scored the ranking of the flipped samples by average precision, AUROC
and precision at $k$, with $k$ the number of flipped samples. A random forest, L2-penalized logistic regression and
$k$-nearest neighbours (correlation distance), each with fixed settings on the 2,000 most variable genes, were scored in
the same way. Contrasts weight the \val{A15} cohort clusters equally, or disease areas equally, with 95\% intervals
from bootstrap resamples of cohort clusters.

\subsection*{Explain}
Explain attributes PRTS's predictions to genes by integrated gradients\cite{sundararajan2017} of the predicted
log-probability, from the context mean in 32 steps, over up to 32 query samples. A gene's importance is its mean
absolute attribution, averaged over the five folds of a task. A gene set's score is the mean absolute per-sample sum
of its genes' attributions, divided by the square root of the set size. Faithfulness was measured by masking. The top
10 genes of a ranking were set to missing in copies of the query samples, and the mean absolute change in PRTS's
prediction was compared between PRTS's top 10 genes and those of moderated $t$\cite{smyth2004,ritchie2015}, and
against 10 random genes. The ratios carry intervals from bootstrap resamples of planted sets or projects. Planted-set
tasks use real TCGA expression of the \val{D13} primary projects, with labels simulated from the standardized mean of
a Hallmark gene set\cite{liberzon2015}; four sets were planted per project (\val{E5} in all), each at two effect
sizes. Chromosome-arm tasks predict a tumour's arm-level copy-number call, with the arm's genes as the truth; the true
arm's rank among all arms' gene sets is given as a percentile (0 is the top).
Disease-module tasks pair a phenotype outside cancer with an established Hallmark programme, such as the interferon
response in systemic lupus erythematosus, and rank the true set among 50 Hallmarks by PRTS's set score, moderated $t$,
preranked GSEA\cite{subramanian2005}, L2 coefficients and LightGBM SHAP values\cite{lundberg2017}.

\subsection*{Evaluation rounds of the applications}
The settings of the four applications were fixed on development cohorts before each held-out run. Held-out evaluation came in two rounds. The
first round used disease-balanced sets of held-out tasks: Audit, \val{A9} tasks; Hub, \val{H20} targets in \val{H21}
task families; Explain, \val{E21} disease-module tasks and 10 TCGA tasks. The confirmation used held-out items that no
first-round run of the same application had touched: Hub, \val{H18} targets in \val{H19} families; Planner, \val{P10} tasks;
Explain, \val{E4} planted-set tasks and \val{E9} chromosome-arm tasks in \val{E10} projects. Between the rounds, the
settings of the Hub, the Planner and Explain were revised on development data, without using any first-round held-out
result; Audit was not revised and was evaluated in the first round only. During the confirmation, the planted-set
Explain results were looked at once after half of their tasks had run. The look was not planned, nothing was changed
after it, and the planned analysis pools both halves.

\subsection*{Use of an AI assistant}

An AI assistant (Claude, Anthropic) assisted with drafting and editing the manuscript. H.W. reviewed and edited the
text and takes responsibility for its content.

\subsection*{Ethics}
This study used publicly available, de-identified data and involved no new recruitment of human participants or
animal experiments.

\subsection*{Author contributions}
H.W. developed the code, conducted the experiments, prepared the figures and wrote the original manuscript.
L.P. reviewed the work, independently reproduced selected experiments and extensively revised the manuscript.
H.L., Z.L., J.H., C.Z. and C.T. reviewed the manuscript and provided research infrastructure and financial support.
All authors approved the manuscript and agreed to its submission.

\subsection*{Funding}
No external grant funding was received. The work was supported by the authors' personal funds and research resources.

\subsection*{Competing interests}
L.P. is employed by Prior Labs, which develops TabPFN. The other authors declare no competing interests.

\subsection*{Acknowledgements}
None.

\subsection*{Data availability}
PreBULK is derived from ARCHS4\cite{lachmann2018}. PreBULK and RealBulk are publicly available at
\url{https://huggingface.co/datasets/hanzi-bulk-collab/bulk-datasets-private}.
The deposited PreBULK is a subsequent quality-controlled edition containing 536,287 profiles; its
\texttt{source\_prebulk\_row} field maps these profiles to the original \val{D14}-profile corpus used for pretraining.
The RealBulk archive contains all \val{D1} curated cohorts, with expression, labels, provenance and harmonized files;
the accompanying \texttt{realbulk-corrections-20261112.zip} overlay supplies the corrected GSE14333 survival labels
and associated metadata.
Cohort lists, held-out task definitions, the clean subset, benchmark scores, robustness tables and source data for
every figure are available at \url{https://github.com/Hanziwww/PRTS/tree/main/tables}.
The same Hugging Face repository provides \texttt{prts-release-data.zip}, containing benchmark expression matrices,
labels, frozen task definitions, per-sample predictions and site-grouped TCGA folds.
TCGA source data are available from the Genomic Data Commons.

\subsection*{Code availability}
The prior that generates the pretraining tasks, the pretraining code, model and inference code, the four applications,
the evaluation pipeline and the post hoc comparator code are publicly available at
\url{https://github.com/Hanziwww/PRTS}. The released training code covers the pretraining procedure.
The evaluated model checkpoint, architecture configuration and checksum are available at
\url{https://huggingface.co/HanZiwww/PRTS}. Author-owned code and model weights are released under Apache License 2.0;
third-party code and source datasets retain their original terms.

\clearpage
\newgeometry{margin=1.5cm}
\nolinenumbers

  \setlegend{fig1}{\figlegend{Fig.~1}{PRTS predicts clinical endpoints from bulk transcriptomes in context.}
\pnl{a}, PRTS was pretrained on 6.0 million prior-generated tasks built from public RNA-seq profiles (Methods). Bars: the final pretraining pass by target branch. PRTS has 220 million parameters and no gene identities. At prediction time it reads labelled context samples and new query samples in one forward pass, and four such passes are averaged. \pnl{b}, The 1,157 usable RealBulk cohorts (193,840 people) by disease area: development and held-out cohorts. \pnl{c}, The held-out benchmarks, one square per task, coloured by task type: RealBulk (175 tasks; 113 within one cohort, 62 across cohorts) and TCGA (44 tasks from 13 projects never used during development). An audit of the split removes 38 RealBulk tasks, leaving a clean set of 137 (Methods). \pnl{d}, The 13 comparators, by family. \pnl{e}, Four applications of the same model, without retraining: label transfer (the Hub), study planning (the Planner), label audit (Audit) and attribution (Explain). Schematics are drawn in code; matrices and curves are illustrative.}
\setlegend{fig2}{\figlegend{Fig.~2}{PRTS has the highest mean accuracy on both held-out benchmarks.}
\pnl{a}, Mean rank correlation between predictions and recorded outcomes (zero, chance; one, a perfect ranking; for binary and survival endpoints, 0.4 equals an AUC or a concordance index of 0.7) of every method on RealBulk (175 tasks) and on TCGA (44 tasks). Top: PRTS and the comparators that predict all four task types, with the task types weighted equally. Bottom: PRTS and the survival-only comparators on the survival tasks. \pnl{b}, PRTS minus each comparator, in the same rows: centres, mean differences; bars, 95\% cluster-bootstrap intervals; open marks, the interval includes zero. Values right of zero favour PRTS. $P$ values, Holm-adjusted over the 13 comparators (link-component bootstrap for RealBulk, exact project-level sign-flip for TCGA), are summarized in Supplementary Fig. 5 and listed in the Source Data. \pnl{c}, Mean rank correlation by task type, with the number of tasks: PRTS (circle), TabPFN v3.5 (triangle), GeneICL (square), TabICLv2 (diamond) and every other comparator (grey ticks). \pnl{d}, Rank of each method by Elo rating in the five views of Supplementary Fig. 7a (L2 in its better configuration per view); PRTS ranks first in every view. The pre-specified analysis, PRTS without class reordering against every comparator, gives the same conclusions (Supplementary Fig. 7b; Methods).}
\setlegend{fig3}{\figlegend{Fig.~3}{Accuracy by setting: few samples, new cohorts and new sites, calibration and cost.}
\pnl{a}, The 43 TCGA tasks large enough for every context size (16 from projects once used to compare checkpoints), with the context subsampled to 32, 64 or 128 people, or in full; mean of the repeats. Shading: between PRTS and the best comparator. \pnl{b}, RealBulk cross-cohort tasks (62) with 0, 5, 10 or 20 labelled target samples added to the context. \pnl{c}, PRTS minus TabPFN v3.5, TabICLv2 and L2 in four settings (centres, mean differences; bars, 95\% cluster-bootstrap intervals; open marks, the interval includes zero; values right of zero favour PRTS). The settings are new patients within one cohort (RealBulk, 113 tasks) and a new cohort (RealBulk across cohorts without target labels, 62 tasks). On TCGA, query patients come from sites present in the context (random folds) or absent from it (site-grouped folds), on the 34 primary-stratum tasks where site-grouped folds are feasible. \pnl{d}, Survival integrated Brier skill score and binary expected calibration error (axis reversed: right is better) on both benchmarks. \pnl{e}, Improvability (mean gap to the best method per task) against time per 1,000 samples; the dashed line is the accuracy--cost front. PRTS is 2.6 times faster than TabPFN v3.5 on RealBulk (37.5 against 96.6 s) and 2.3 times faster on TCGA. Every comparator's curves in \pnl{a} and \pnl{b} are in Supplementary Fig. 8, and every probabilistic score in Supplementary Fig. 9c.}
\setlegend{fig4}{\figlegend{Fig.~4}{Label transfer: the Hub labels a new cohort from a hub of public reference cohorts.}
\pnl{a}, For each endpoint the Hub keeps a hub of labelled public reference cohorts (one task family). They are the context; the unlabelled target cohort is the query. The entropy of the predictions tracks how well a transfer works. \pnl{b}, Hub minus each comparator (rank correlation; task families weighted equally; 95\% family-bootstrap intervals; open marks include zero; values right of zero favour the Hub): confirmation, 73 targets in 47 families; first round, 45 targets in 33 families. TabPFN v3.5 and GeneICL read the Hub's own reference draws. The SingleR-style rows use the targets with classes (48 confirmation targets in 33 families; 34 first-round targets in 24 families). \pnl{c}, Entropy against realized rank correlation per target with classes (48 confirmation targets, Spearman $\rho$ = \textminus{}0.79; 34 first-round targets, $\rho$ = \textminus{}0.66). \pnl{d}, Three published signatures on four held-out cohorts: the locked published score (open) and the Hub (filled), with Hub minus published score (95\% patient-bootstrap interval, which excludes zero for GSE116918 only). Top to bottom: cell-cycle progression score in prostate cancer, GSE116918; interferon score in lupus, GSE154851; cell-cycle progression score, GSE220095; three-gene tuberculosis signature, GSE73408. \pnl{e}, Confirmation by task type.}
\setlegend{fig5}{\figlegend{Fig.~5}{Study planning: the Planner forecasts, from a 30-person pilot, the accuracy PRTS reaches on a full cohort.}
\pnl{a}, A pilot's learning curve and a prior per task type, learned on development cohorts, give a forecast at the full size with an 80\% interval. \pnl{b}, Forecast against realized rank correlation at the full size (64 held-out tasks; mean over five pilots per task); the 80\% intervals, drawn up to 1, the largest possible rank correlation, cover 75\% of outcomes. \pnl{c}, Mean absolute error of the forecast over the 64 tasks: the Planner, a pooled prior, least-squares extrapolation and the pilot's own value. Planner minus least squares: \textminus{}0.080 (95\% CI \textminus{}0.108 to \textminus{}0.057); error ratio 0.69. \pnl{d}, Mean absolute error by task type and by disease area (number of tasks). \pnl{e}, Feasibility test (19 label permutations): $P$ values with shuffled labels (size; 1 of 40 tasks rejected) and with real labels (power; 16 of 20 rejected). Shading: rejected at $P$ $\leq$ 0.05, that is, the task is learnable.}
\setlegend{fig6}{\figlegend{Fig.~6}{Label audit and attribution: Audit ranks likely mislabelled samples; Explain finds the genes behind predictions.}
\pnl{a}, Audit ranks a cohort's samples by how likely their label is wrong. \pnl{b}, Audit's average precision, AUROC and precision at k relative to a random forest, with L2 and k-nearest neighbours (47 held-out tasks in which 10\% of the labels were flipped at random). \pnl{c}, Audit minus random forest, average precision, with cohort clusters weighted equally and with disease areas weighted equally (95\% intervals; values right of zero favour Audit). \pnl{d}, Explain's faithfulness: masking PRTS's top 10 genes changes PRTS's predictions more than masking moderated t's top 10 (fold change; planted sets, 104 tasks; TCGA chromosome-arm tasks, 22). \pnl{e}, Change from masking PRTS's top 10 genes relative to masking 10 random genes (first round, descriptive). \pnl{f}, Rank percentile of the true chromosome arm on TCGA: PRTS and moderated t (difference \textminus{}0.018; 95\% CI \textminus{}0.048 to \textminus{}0.001). \pnl{g}, Mean rank of the true Hallmark programme among 50 (21 tasks, first round, descriptive).}
\setlegend{ed01}{\figlegend{Supplementary Fig.~1}{The data: PreBULK for the prior, RealBulk for development and evaluation, TCGA for evaluation.}
\pnl{a}, Provenance. PreBULK (581,503 ARCHS4-derived human RNA-seq profiles, 20,010 genes, ln(TPM + 1)) passes the prior's quality control (transcriptomic libraries, zero fraction, study size, duplicate profiles) and yields its assets (gene quantiles, co-expression modules, principal components). RealBulk's 1,160 cohorts were screened (at least 50 labelled people per endpoint; *at least 30 in a relaxed tier of 57 development cohorts), curated (labels verbatim, one sample per person) and deduplicated (sample-level overlap scan on gene ranks); link components were assigned to development, held-out or unusable cohorts. No benchmark cohort has its own samples in PreBULK; the audit of the split found four held-out RNA-seq cohorts that do, none with a benchmark task (Methods). The TCGA panel comes from GDC open-access data; a content scan removed 3 patients resembling PreBULK profiles. \pnl{b}, Usable cohorts by disease group (tile area proportional to cohorts), coloured by disease area; CRC, colorectal cancer; TB, tuberculosis; RA, rheumatoid arthritis; UC, ulcerative colitis; MS, multiple sclerosis. \pnl{c}, People per cohort. \pnl{d}, Platform by cancer and other diseases (cohorts). \pnl{e}, Endpoint types per cohort (UpSet plot).}
\setlegend{ed02}{\figlegend{Supplementary Fig.~2}{How tasks are sampled from public expression profiles and anonymized.}
\pnl{a}, The features x of a task are sampled from PreBULK (samples from one or more studies; a subset of genes chosen by one of five rules). Tasks from RealBulk's development cohorts are sampled in the same way (people of a cohort; a subset of the genes it measures) after harmonization onto PreBULK's gene universe and log scale. For PreBULK tasks, one of four branches makes the target (recorded sample annotations; a masked gene, gene-set or principal-component score; a causal graph rooted in real genes; latent programmes mapped onto real gene marginals through 64-point quantile tables). A head then gives it a task type, and every value distribution is re-expressed with parameters drawn per task (19 value formats, sequencing depth, dispersion, warps and batch shifts). Every task is anonymized: gene order is shuffled, samples and class labels are re-ordered, no gene names are kept, and the model sees random gene codes only. Four gates admit a prior task (validity, learnability, discoverability from a small context, a difficulty quota). Ribbon widths are the final pretraining pass's shares by branch and by task type. \pnl{b}, Generated tasks (illustrative): context samples above, query samples below, the target column on the right. \pnl{c}, Pretraining volume in three passes of increasing width (6.04 million tasks).}
\setlegend{ed03}{\figlegend{Supplementary Fig.~3}{The PRTS architecture.}
\pnl{a}, The forward pass: a cell tokenizer; four cell blocks, each with attention along samples, a feed-forward layer and attention along genes, every sub-layer with a residual sum; row compression; twenty row blocks (attention across samples and a feed-forward layer); a readout; and three heads. \pnl{b}, A cell token (192 dimensions) is the sum of the expression value's encoding (the value and 16 radial bases), a random gene code (re-drawn every pass), the target of a context sample or an 'unknown' token for a query sample, and the task type. \pnl{c}, Attention along samples, within one gene: every sample reads the context samples (and eight gene summaries) only. \pnl{d}, Attention along genes, within one sample, through 128 inducing tokens and 4 routing tokens: the cost grows linearly with the number of genes. \pnl{e}, Eight learned summaries of 128 dimensions read each sample's cells into one 1,024-dimensional row; query rows read context rows. \pnl{f}, Inference: four forward passes with their own random gene codes and, for classes, their own class order, averaged. \pnl{g}, The 219,595,334 parameters by component. \pnl{h}, Class probabilities (up to 64 classes), a predictive distribution from 999 quantiles and a mixture of eight log-normals for survival. Drawn in code from the model definition; nodes and curves are illustrative.}
\setlegend{ed04}{\figlegend{Supplementary Fig.~4}{Robustness of the comparison: the clean held-out set, TCGA site structure and post hoc comparators.}
\pnl{a}, PRTS minus each comparator of the primary family on all 175 RealBulk tasks and on the 137 clean tasks, which touch no cohort that the audit of the split linked to a development cohort (bootstrap over link components, rebuilt ones on the clean set; values right of zero favour PRTS, here and in \pnl{c}). \pnl{b}, Top: rank correlation of a site-only predictor (the context's mean label at each patient's tissue-source site; random folds), one tick per TCGA task, by stratum; diamonds, medians. Bottom: each method's change in rank correlation from random folds (known sites) to site-grouped folds (new sites) on the 34 primary-stratum tasks where site-grouped folds are feasible (bootstrap over projects). \pnl{c}, PRTS minus each post hoc comparator, configured on the development tuning tasks, on RealBulk (all tasks, within one cohort, across cohorts) and on TCGA (44 tasks). Survival-only and classification-only comparators are scored on their task types; a dash marks a comparator not run on that benchmark. Holm-adjusted $P$ values among the post hoc comparators are in Supplementary Table 3.}
\setlegend{ed05}{\figlegend{Supplementary Fig.~5}{Rank correlation of every method by task type, and PRTS minus each comparator.}
\pnl{a}, Mean rank correlation by task type and over all tasks on both benchmarks; bottom row, PRTS minus the strongest comparator in each column, negative where a comparator is higher. \pnl{b}, PRTS minus each comparator with 95\% intervals, by task type (descriptive) and over all tasks. The overall contrasts have Holm-adjusted $P$ values of at most 0.016 on RealBulk, except DeepSurv (0.21), and of at most 0.031 on TCGA.}
\setlegend{ed06}{\figlegend{Supplementary Fig.~6}{Head to head.}
\pnl{a}, Share of tasks on which PRTS's rank correlation is higher than (wins), within 0.01 of (ties) or lower than (losses) each comparator's. \pnl{b}, Share of tasks on which each method ranks first among PRTS and the 13 comparators (tied methods each count). \pnl{c}, Rank correlation of PRTS against TabPFN v3.5, one point per task, coloured by task type; counts use the 0.01 tie margin.}
\setlegend{ed07}{\figlegend{Supplementary Fig.~7}{Elo ratings, the pre-specified analysis and the cost of class reordering.}
\pnl{a}, Bradley--Terry (Elo) ratings from per-task games with 95\% bootstrap intervals in five views: all tasks of both panels (descriptive), RealBulk, RealBulk across cohorts, TCGA projects never used in development and TCGA projects once used to compare checkpoints. L2 is rated in its better configuration in each view (all genes on TCGA projects once used to compare checkpoints, 2,000 genes elsewhere). \pnl{b}, The pre-specified analysis: PRTS without class reordering, the pre-specified arm, minus each comparator; $P$ values Holm-adjusted over the 13 comparators, among them GeneICL, which is outside the protocol (RealBulk, link-component bootstrap; TCGA, exact project-level sign-flip, whose $P$ values take discrete values). \pnl{c}, GPU seconds for the main design with and without class reordering.}
\setlegend{ed08}{\figlegend{Supplementary Fig.~8}{Few samples, few target labels and cost, against every comparator.}
\pnl{a}, The 43 TCGA tasks large enough for every context size (16 from projects once used to compare checkpoints), with the context subsampled to 32, 64 or 128 people, or in full: PRTS (blue) and each comparator; shading, between PRTS and the best comparator. \pnl{b}, RealBulk cross-cohort tasks with 0, 5, 10 or 20 labelled target samples. \pnl{c}, Improvability against time per 1,000 samples in four views; dashed, the accuracy--cost front. Times are those of the first runs; the corrected cross-cohort runs ran on an oversubscribed processor.}
\setlegend{ed09}{\figlegend{Supplementary Fig.~9}{Strata and probabilistic scores.}
\pnl{a}, Mean rank correlation within the levels of four strata: cohort setting; cancer or other diseases; whether the target concept occurs among the development cohorts; TCGA projects never used in development or once used to compare checkpoints. \pnl{b}, PRTS minus TabPFN v3.5 and minus GeneICL in each level (95\% intervals, descriptive). \pnl{c}, Every probabilistic score on both benchmarks: log-loss and Brier skill scores (one minus the geometric mean ratio of the score to the null model's) and expected calibration error (axis reversed: right is better) for binary and multiclass tasks, the quantile skill score for continuous tasks and the integrated Brier skill score for survival tasks.}

  \newcommand{\figpage}[2]{\begin{center}\includegraphics[width=180mm,height=175mm,keepaspectratio]{#1.pdf}
    \end{center}{\small\setstretch{1}\uselegend{#1}\par}\clearpage}

\section*{Figures}
\figpage{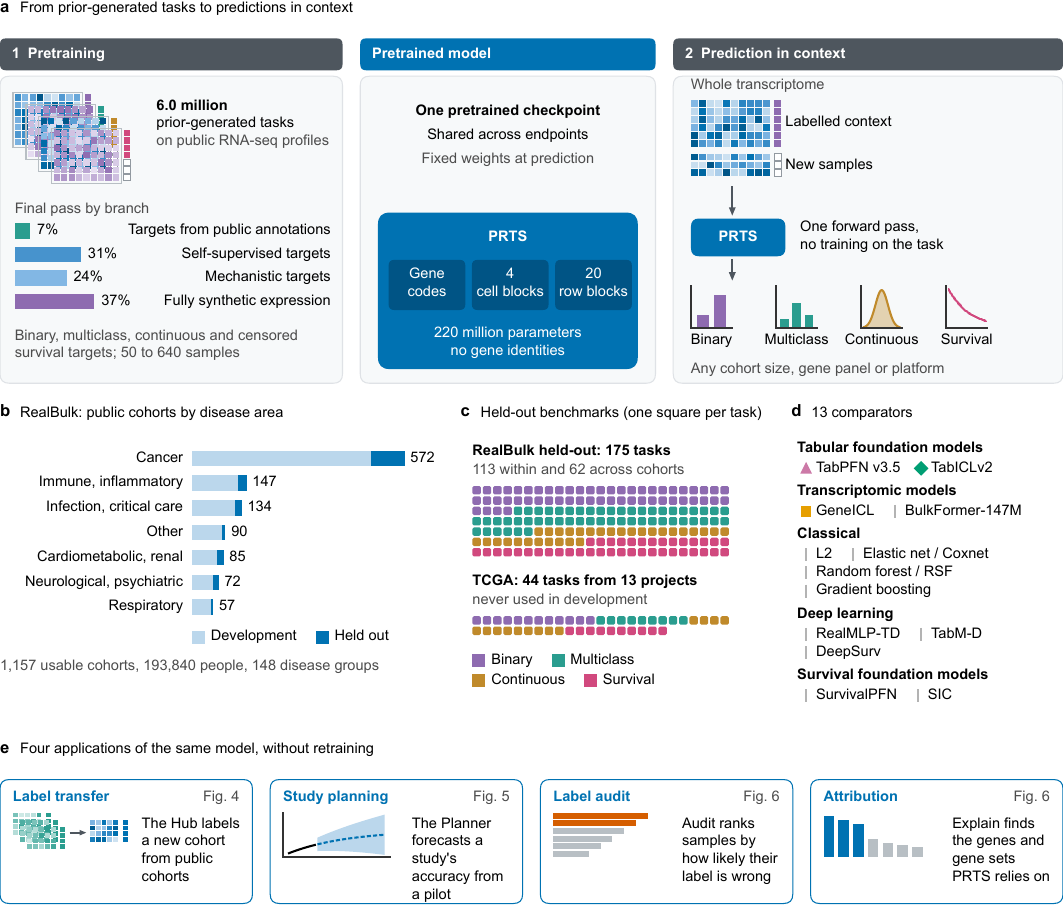}{Fig.~1}
\figpage{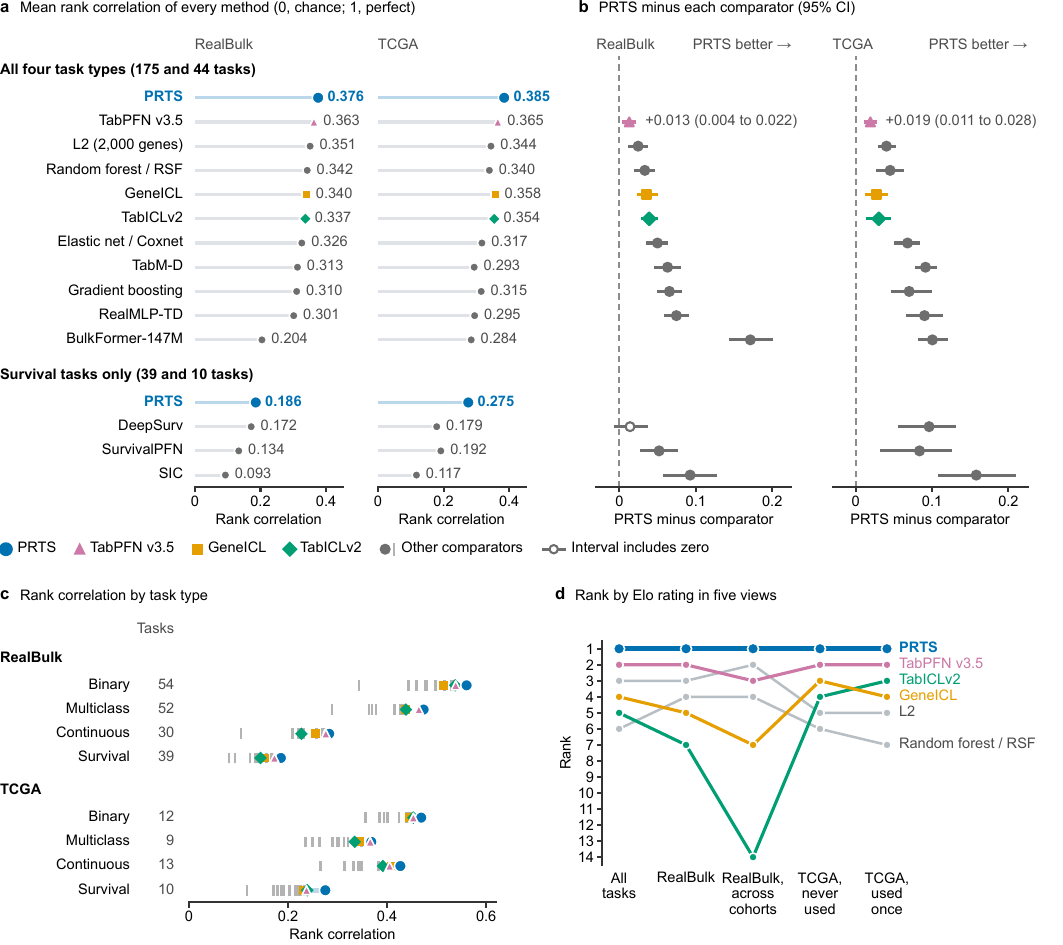}{Fig.~2}
\figpage{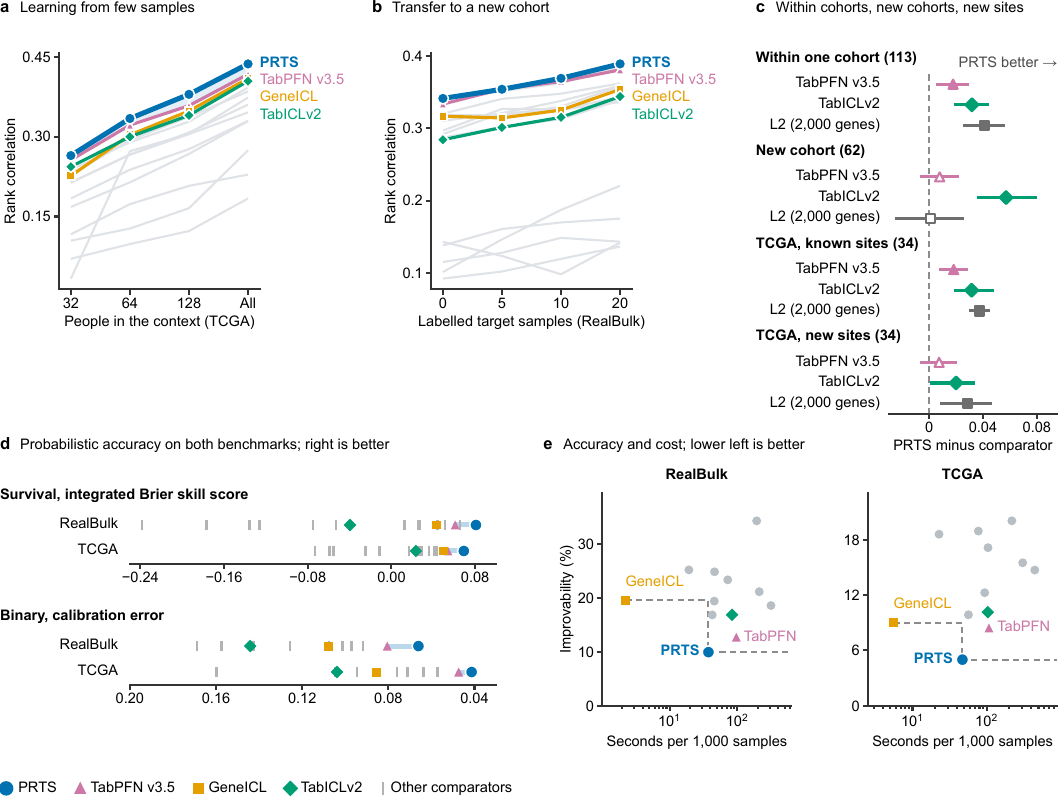}{Fig.~3}
\figpage{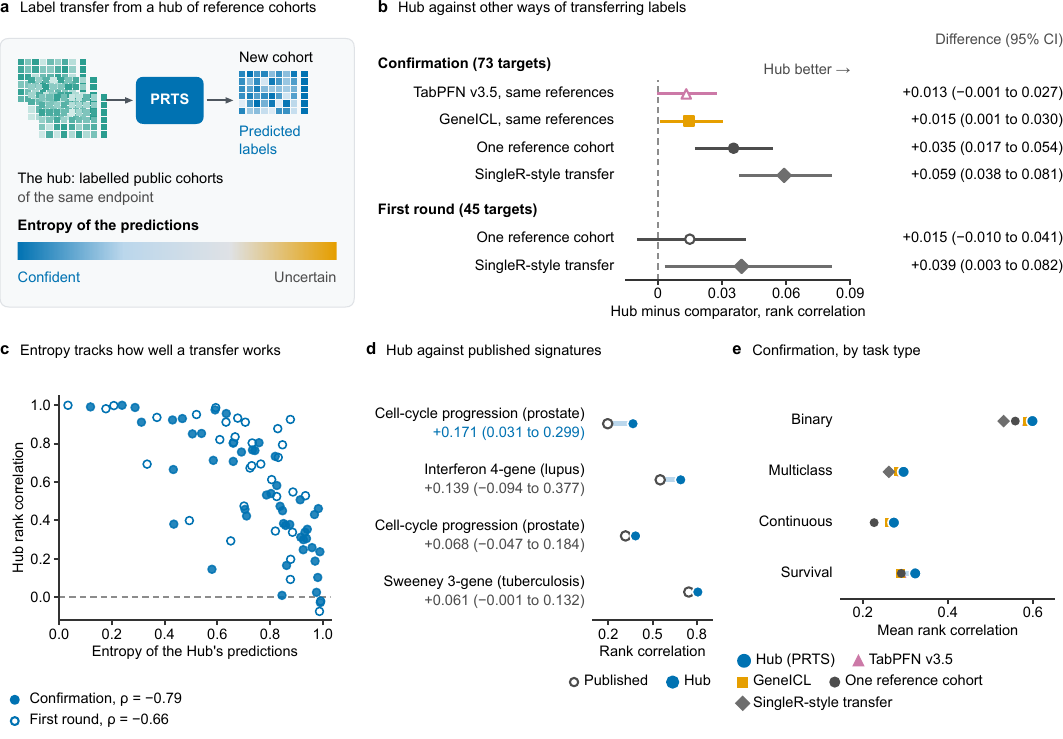}{Fig.~4}
\figpage{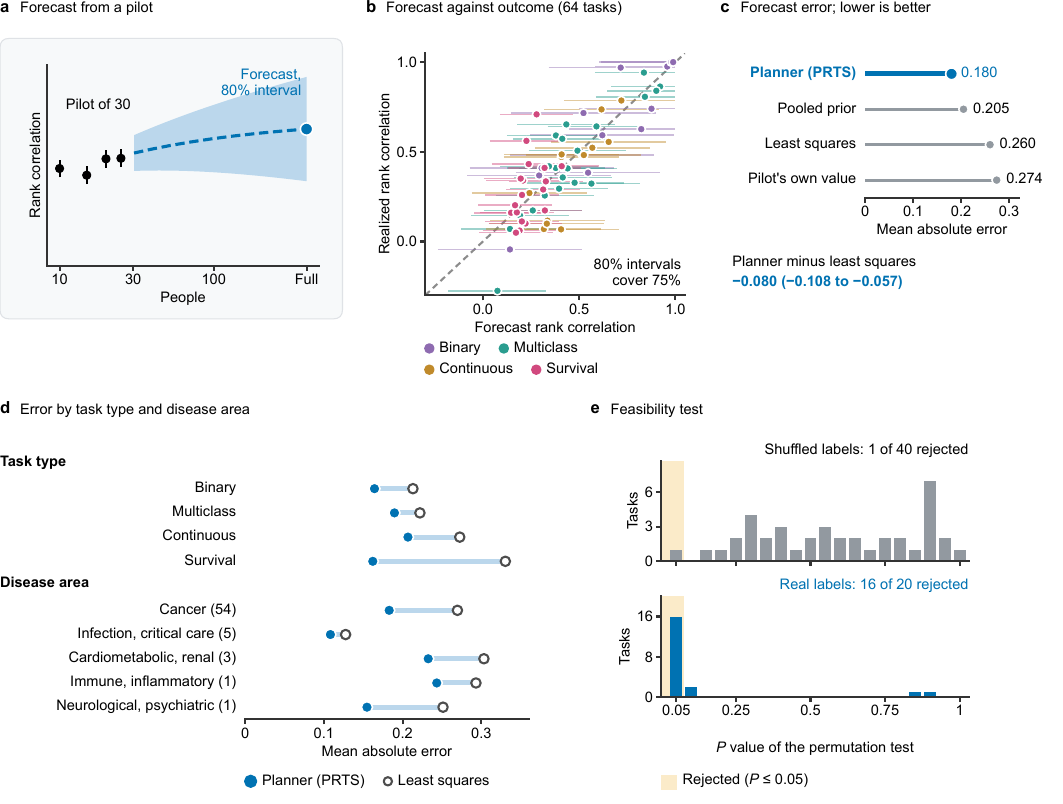}{Fig.~5}
\figpage{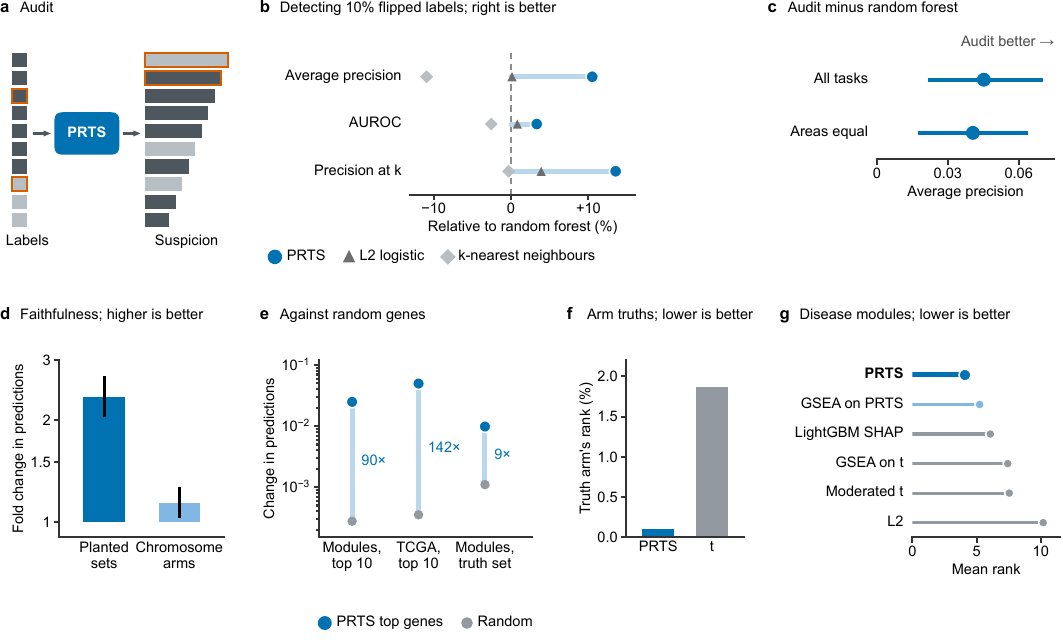}{Fig.~6}
\restoregeometry

\clearpage
\setstretch{1.1}
\renewcommand{\tablename}{Supplementary Table}
\setlength{\LTcapwidth}{\textwidth}

\begin{center}
{\Large\bfseries Supplementary Information\par}
\medskip
{\large \papertitle\par}
\bigskip
{Han~Wang\textsuperscript{1,*}, Lennart~Purucker\textsuperscript{2},
Huafeng~Liu\textsuperscript{3}, Ziyin~Luo\textsuperscript{4,5},
Junfeng~Hao\textsuperscript{6,7}, Chao~Zhu\textsuperscript{8,9},
Chunjie~Tian\textsuperscript{1,6,10,*}\par}
\end{center}

\begingroup
\small
\setstretch{1.0}
\setlength{\parindent}{0pt}
\setlength{\parskip}{4pt}
\textsuperscript{1}Department of Otorhinolaryngology, Affiliated Hospital of Guangdong Medical University, Zhanjiang 524000, Guangdong, China.\par
\textsuperscript{2}Prior Labs, Freiburg im Breisgau 79098, Germany.\par
\textsuperscript{3}Guangdong Provincial Key Laboratory of Autophagy and Major Chronic Non-communicable Diseases, Zhanjiang 524001, Guangdong, China.\par
\textsuperscript{4}Department of Otorhinolaryngology Head and Neck Surgery, The Sixth Affiliated Hospital, Sun Yat-sen University, Guangzhou 510655, China.\par
\textsuperscript{5}Department of Medicine, University of Pennsylvania, Philadelphia, Pennsylvania, USA.\par
\textsuperscript{6}Guangdong Provincial Key Laboratory of Autophagy and Major Chronic Non-communicable Diseases, Key Laboratory of Prevention and Management of Chronic Kidney Disease of Zhanjiang City, Institute of Nephrology, Affiliated Hospital of Guangdong Medical University, Zhanjiang 524001, Guangdong, China.\par
\textsuperscript{7}Department of Family Medicine, Shengjing Hospital of China Medical University, Shenyang 110022, Liaoning, China.\par
\textsuperscript{8}Department of Rheumatology and Immunology, The Third Affiliated Hospital of Naval Medical University, Shanghai 201805, China.\par
\textsuperscript{9}Department of Tropical Diseases, Naval Medical University, Shanghai 200433, China.\par
\textsuperscript{10}Otolaryngology Research Center, Guangdong Medical University Affiliated Hospital, Zhanjiang 524000, Guangdong, China.\par
\medskip
\textsuperscript{*}Correspondence: Han Wang, \href{mailto:han@hanziwww.com}{\nolinkurl{han@hanziwww.com}}; Chunjie Tian, \href{mailto:tianchunjie@gdmu.edu.cn}{\nolinkurl{tianchunjie@gdmu.edu.cn}}.\par
\endgroup
\bigskip

\noindent\textbf{Contents.} Supplementary Figs~1--9. Supplementary Note~1: the corrected gene ranking of the comparators and both versions of the
affected results. Supplementary Note~2: the audit of the split between development and held-out cohorts.
Supplementary Note~3: post hoc attempts to improve PRTS. Supplementary Tables~1--7.

\clearpage
\newgeometry{margin=1.5cm}
\setlegend{fig1}{\figlegend{Fig.~1}{PRTS predicts clinical endpoints from bulk transcriptomes in context.}
\pnl{a}, PRTS was pretrained on 6.0 million prior-generated tasks built from public RNA-seq profiles (Methods). Bars: the final pretraining pass by target branch. PRTS has 220 million parameters and no gene identities. At prediction time it reads labelled context samples and new query samples in one forward pass, and four such passes are averaged. \pnl{b}, The 1,157 usable RealBulk cohorts (193,840 people) by disease area: development and held-out cohorts. \pnl{c}, The held-out benchmarks, one square per task, coloured by task type: RealBulk (175 tasks; 113 within one cohort, 62 across cohorts) and TCGA (44 tasks from 13 projects never used during development). An audit of the split removes 38 RealBulk tasks, leaving a clean set of 137 (Methods). \pnl{d}, The 13 comparators, by family. \pnl{e}, Four applications of the same model, without retraining: label transfer (the Hub), study planning (the Planner), label audit (Audit) and attribution (Explain). Schematics are drawn in code; matrices and curves are illustrative.}
\setlegend{fig2}{\figlegend{Fig.~2}{PRTS has the highest mean accuracy on both held-out benchmarks.}
\pnl{a}, Mean rank correlation between predictions and recorded outcomes (zero, chance; one, a perfect ranking; for binary and survival endpoints, 0.4 equals an AUC or a concordance index of 0.7) of every method on RealBulk (175 tasks) and on TCGA (44 tasks). Top: PRTS and the comparators that predict all four task types, with the task types weighted equally. Bottom: PRTS and the survival-only comparators on the survival tasks. \pnl{b}, PRTS minus each comparator, in the same rows: centres, mean differences; bars, 95\% cluster-bootstrap intervals; open marks, the interval includes zero. Values right of zero favour PRTS. $P$ values, Holm-adjusted over the 13 comparators (link-component bootstrap for RealBulk, exact project-level sign-flip for TCGA), are summarized in Supplementary Fig. 5 and listed in the Source Data. \pnl{c}, Mean rank correlation by task type, with the number of tasks: PRTS (circle), TabPFN v3.5 (triangle), GeneICL (square), TabICLv2 (diamond) and every other comparator (grey ticks). \pnl{d}, Rank of each method by Elo rating in the five views of Supplementary Fig. 7a (L2 in its better configuration per view); PRTS ranks first in every view. The pre-specified analysis, PRTS without class reordering against every comparator, gives the same conclusions (Supplementary Fig. 7b; Methods).}
\setlegend{fig3}{\figlegend{Fig.~3}{Accuracy by setting: few samples, new cohorts and new sites, calibration and cost.}
\pnl{a}, The 43 TCGA tasks large enough for every context size (16 from projects once used to compare checkpoints), with the context subsampled to 32, 64 or 128 people, or in full; mean of the repeats. Shading: between PRTS and the best comparator. \pnl{b}, RealBulk cross-cohort tasks (62) with 0, 5, 10 or 20 labelled target samples added to the context. \pnl{c}, PRTS minus TabPFN v3.5, TabICLv2 and L2 in four settings (centres, mean differences; bars, 95\% cluster-bootstrap intervals; open marks, the interval includes zero; values right of zero favour PRTS). The settings are new patients within one cohort (RealBulk, 113 tasks) and a new cohort (RealBulk across cohorts without target labels, 62 tasks). On TCGA, query patients come from sites present in the context (random folds) or absent from it (site-grouped folds), on the 34 primary-stratum tasks where site-grouped folds are feasible. \pnl{d}, Survival integrated Brier skill score and binary expected calibration error (axis reversed: right is better) on both benchmarks. \pnl{e}, Improvability (mean gap to the best method per task) against time per 1,000 samples; the dashed line is the accuracy--cost front. PRTS is 2.6 times faster than TabPFN v3.5 on RealBulk (37.5 against 96.6 s) and 2.3 times faster on TCGA. Every comparator's curves in \pnl{a} and \pnl{b} are in Supplementary Fig. 8, and every probabilistic score in Supplementary Fig. 9c.}
\setlegend{fig4}{\figlegend{Fig.~4}{Label transfer: the Hub labels a new cohort from a hub of public reference cohorts.}
\pnl{a}, For each endpoint the Hub keeps a hub of labelled public reference cohorts (one task family). They are the context; the unlabelled target cohort is the query. The entropy of the predictions tracks how well a transfer works. \pnl{b}, Hub minus each comparator (rank correlation; task families weighted equally; 95\% family-bootstrap intervals; open marks include zero; values right of zero favour the Hub): confirmation, 73 targets in 47 families; first round, 45 targets in 33 families. TabPFN v3.5 and GeneICL read the Hub's own reference draws. The SingleR-style rows use the targets with classes (48 confirmation targets in 33 families; 34 first-round targets in 24 families). \pnl{c}, Entropy against realized rank correlation per target with classes (48 confirmation targets, Spearman $\rho$ = \textminus{}0.79; 34 first-round targets, $\rho$ = \textminus{}0.66). \pnl{d}, Three published signatures on four held-out cohorts: the locked published score (open) and the Hub (filled), with Hub minus published score (95\% patient-bootstrap interval, which excludes zero for GSE116918 only). Top to bottom: cell-cycle progression score in prostate cancer, GSE116918; interferon score in lupus, GSE154851; cell-cycle progression score, GSE220095; three-gene tuberculosis signature, GSE73408. \pnl{e}, Confirmation by task type.}
\setlegend{fig5}{\figlegend{Fig.~5}{Study planning: the Planner forecasts, from a 30-person pilot, the accuracy PRTS reaches on a full cohort.}
\pnl{a}, A pilot's learning curve and a prior per task type, learned on development cohorts, give a forecast at the full size with an 80\% interval. \pnl{b}, Forecast against realized rank correlation at the full size (64 held-out tasks; mean over five pilots per task); the 80\% intervals, drawn up to 1, the largest possible rank correlation, cover 75\% of outcomes. \pnl{c}, Mean absolute error of the forecast over the 64 tasks: the Planner, a pooled prior, least-squares extrapolation and the pilot's own value. Planner minus least squares: \textminus{}0.080 (95\% CI \textminus{}0.108 to \textminus{}0.057); error ratio 0.69. \pnl{d}, Mean absolute error by task type and by disease area (number of tasks). \pnl{e}, Feasibility test (19 label permutations): $P$ values with shuffled labels (size; 1 of 40 tasks rejected) and with real labels (power; 16 of 20 rejected). Shading: rejected at $P$ $\leq$ 0.05, that is, the task is learnable.}
\setlegend{fig6}{\figlegend{Fig.~6}{Label audit and attribution: Audit ranks likely mislabelled samples; Explain finds the genes behind predictions.}
\pnl{a}, Audit ranks a cohort's samples by how likely their label is wrong. \pnl{b}, Audit's average precision, AUROC and precision at k relative to a random forest, with L2 and k-nearest neighbours (47 held-out tasks in which 10\% of the labels were flipped at random). \pnl{c}, Audit minus random forest, average precision, with cohort clusters weighted equally and with disease areas weighted equally (95\% intervals; values right of zero favour Audit). \pnl{d}, Explain's faithfulness: masking PRTS's top 10 genes changes PRTS's predictions more than masking moderated t's top 10 (fold change; planted sets, 104 tasks; TCGA chromosome-arm tasks, 22). \pnl{e}, Change from masking PRTS's top 10 genes relative to masking 10 random genes (first round, descriptive). \pnl{f}, Rank percentile of the true chromosome arm on TCGA: PRTS and moderated t (difference \textminus{}0.018; 95\% CI \textminus{}0.048 to \textminus{}0.001). \pnl{g}, Mean rank of the true Hallmark programme among 50 (21 tasks, first round, descriptive).}
\setlegend{ed01}{\figlegend{Supplementary Fig.~1}{The data: PreBULK for the prior, RealBulk for development and evaluation, TCGA for evaluation.}
\pnl{a}, Provenance. PreBULK (581,503 ARCHS4-derived human RNA-seq profiles, 20,010 genes, ln(TPM + 1)) passes the prior's quality control (transcriptomic libraries, zero fraction, study size, duplicate profiles) and yields its assets (gene quantiles, co-expression modules, principal components). RealBulk's 1,160 cohorts were screened (at least 50 labelled people per endpoint; *at least 30 in a relaxed tier of 57 development cohorts), curated (labels verbatim, one sample per person) and deduplicated (sample-level overlap scan on gene ranks); link components were assigned to development, held-out or unusable cohorts. No benchmark cohort has its own samples in PreBULK; the audit of the split found four held-out RNA-seq cohorts that do, none with a benchmark task (Methods). The TCGA panel comes from GDC open-access data; a content scan removed 3 patients resembling PreBULK profiles. \pnl{b}, Usable cohorts by disease group (tile area proportional to cohorts), coloured by disease area; CRC, colorectal cancer; TB, tuberculosis; RA, rheumatoid arthritis; UC, ulcerative colitis; MS, multiple sclerosis. \pnl{c}, People per cohort. \pnl{d}, Platform by cancer and other diseases (cohorts). \pnl{e}, Endpoint types per cohort (UpSet plot).}
\setlegend{ed02}{\figlegend{Supplementary Fig.~2}{How tasks are sampled from public expression profiles and anonymized.}
\pnl{a}, The features x of a task are sampled from PreBULK (samples from one or more studies; a subset of genes chosen by one of five rules). Tasks from RealBulk's development cohorts are sampled in the same way (people of a cohort; a subset of the genes it measures) after harmonization onto PreBULK's gene universe and log scale. For PreBULK tasks, one of four branches makes the target (recorded sample annotations; a masked gene, gene-set or principal-component score; a causal graph rooted in real genes; latent programmes mapped onto real gene marginals through 64-point quantile tables). A head then gives it a task type, and every value distribution is re-expressed with parameters drawn per task (19 value formats, sequencing depth, dispersion, warps and batch shifts). Every task is anonymized: gene order is shuffled, samples and class labels are re-ordered, no gene names are kept, and the model sees random gene codes only. Four gates admit a prior task (validity, learnability, discoverability from a small context, a difficulty quota). Ribbon widths are the final pretraining pass's shares by branch and by task type. \pnl{b}, Generated tasks (illustrative): context samples above, query samples below, the target column on the right. \pnl{c}, Pretraining volume in three passes of increasing width (6.04 million tasks).}
\setlegend{ed03}{\figlegend{Supplementary Fig.~3}{The PRTS architecture.}
\pnl{a}, The forward pass: a cell tokenizer; four cell blocks, each with attention along samples, a feed-forward layer and attention along genes, every sub-layer with a residual sum; row compression; twenty row blocks (attention across samples and a feed-forward layer); a readout; and three heads. \pnl{b}, A cell token (192 dimensions) is the sum of the expression value's encoding (the value and 16 radial bases), a random gene code (re-drawn every pass), the target of a context sample or an 'unknown' token for a query sample, and the task type. \pnl{c}, Attention along samples, within one gene: every sample reads the context samples (and eight gene summaries) only. \pnl{d}, Attention along genes, within one sample, through 128 inducing tokens and 4 routing tokens: the cost grows linearly with the number of genes. \pnl{e}, Eight learned summaries of 128 dimensions read each sample's cells into one 1,024-dimensional row; query rows read context rows. \pnl{f}, Inference: four forward passes with their own random gene codes and, for classes, their own class order, averaged. \pnl{g}, The 219,595,334 parameters by component. \pnl{h}, Class probabilities (up to 64 classes), a predictive distribution from 999 quantiles and a mixture of eight log-normals for survival. Drawn in code from the model definition; nodes and curves are illustrative.}
\setlegend{ed04}{\figlegend{Supplementary Fig.~4}{Robustness of the comparison: the clean held-out set, TCGA site structure and post hoc comparators.}
\pnl{a}, PRTS minus each comparator of the primary family on all 175 RealBulk tasks and on the 137 clean tasks, which touch no cohort that the audit of the split linked to a development cohort (bootstrap over link components, rebuilt ones on the clean set; values right of zero favour PRTS, here and in \pnl{c}). \pnl{b}, Top: rank correlation of a site-only predictor (the context's mean label at each patient's tissue-source site; random folds), one tick per TCGA task, by stratum; diamonds, medians. Bottom: each method's change in rank correlation from random folds (known sites) to site-grouped folds (new sites) on the 34 primary-stratum tasks where site-grouped folds are feasible (bootstrap over projects). \pnl{c}, PRTS minus each post hoc comparator, configured on the development tuning tasks, on RealBulk (all tasks, within one cohort, across cohorts) and on TCGA (44 tasks). Survival-only and classification-only comparators are scored on their task types; a dash marks a comparator not run on that benchmark. Holm-adjusted $P$ values among the post hoc comparators are in Supplementary Table 3.}
\setlegend{ed05}{\figlegend{Supplementary Fig.~5}{Rank correlation of every method by task type, and PRTS minus each comparator.}
\pnl{a}, Mean rank correlation by task type and over all tasks on both benchmarks; bottom row, PRTS minus the strongest comparator in each column, negative where a comparator is higher. \pnl{b}, PRTS minus each comparator with 95\% intervals, by task type (descriptive) and over all tasks. The overall contrasts have Holm-adjusted $P$ values of at most 0.016 on RealBulk, except DeepSurv (0.21), and of at most 0.031 on TCGA.}
\setlegend{ed06}{\figlegend{Supplementary Fig.~6}{Head to head.}
\pnl{a}, Share of tasks on which PRTS's rank correlation is higher than (wins), within 0.01 of (ties) or lower than (losses) each comparator's. \pnl{b}, Share of tasks on which each method ranks first among PRTS and the 13 comparators (tied methods each count). \pnl{c}, Rank correlation of PRTS against TabPFN v3.5, one point per task, coloured by task type; counts use the 0.01 tie margin.}
\setlegend{ed07}{\figlegend{Supplementary Fig.~7}{Elo ratings, the pre-specified analysis and the cost of class reordering.}
\pnl{a}, Bradley--Terry (Elo) ratings from per-task games with 95\% bootstrap intervals in five views: all tasks of both panels (descriptive), RealBulk, RealBulk across cohorts, TCGA projects never used in development and TCGA projects once used to compare checkpoints. L2 is rated in its better configuration in each view (all genes on TCGA projects once used to compare checkpoints, 2,000 genes elsewhere). \pnl{b}, The pre-specified analysis: PRTS without class reordering, the pre-specified arm, minus each comparator; $P$ values Holm-adjusted over the 13 comparators, among them GeneICL, which is outside the protocol (RealBulk, link-component bootstrap; TCGA, exact project-level sign-flip, whose $P$ values take discrete values). \pnl{c}, GPU seconds for the main design with and without class reordering.}
\setlegend{ed08}{\figlegend{Supplementary Fig.~8}{Few samples, few target labels and cost, against every comparator.}
\pnl{a}, The 43 TCGA tasks large enough for every context size (16 from projects once used to compare checkpoints), with the context subsampled to 32, 64 or 128 people, or in full: PRTS (blue) and each comparator; shading, between PRTS and the best comparator. \pnl{b}, RealBulk cross-cohort tasks with 0, 5, 10 or 20 labelled target samples. \pnl{c}, Improvability against time per 1,000 samples in four views; dashed, the accuracy--cost front. Times are those of the first runs; the corrected cross-cohort runs ran on an oversubscribed processor.}
\setlegend{ed09}{\figlegend{Supplementary Fig.~9}{Strata and probabilistic scores.}
\pnl{a}, Mean rank correlation within the levels of four strata: cohort setting; cancer or other diseases; whether the target concept occurs among the development cohorts; TCGA projects never used in development or once used to compare checkpoints. \pnl{b}, PRTS minus TabPFN v3.5 and minus GeneICL in each level (95\% intervals, descriptive). \pnl{c}, Every probabilistic score on both benchmarks: log-loss and Brier skill scores (one minus the geometric mean ratio of the score to the null model's) and expected calibration error (axis reversed: right is better) for binary and multiclass tasks, the quantile skill score for continuous tasks and the integrated Brier skill score for survival tasks.}

\newcommand{\suppfigpage}[1]{\begin{center}\includegraphics[width=180mm,height=175mm,keepaspectratio]{#1.pdf}
  \end{center}{\small\setstretch{1}\uselegend{#1}\par}\clearpage}
\section*{Supplementary Figures}
\suppfigpage{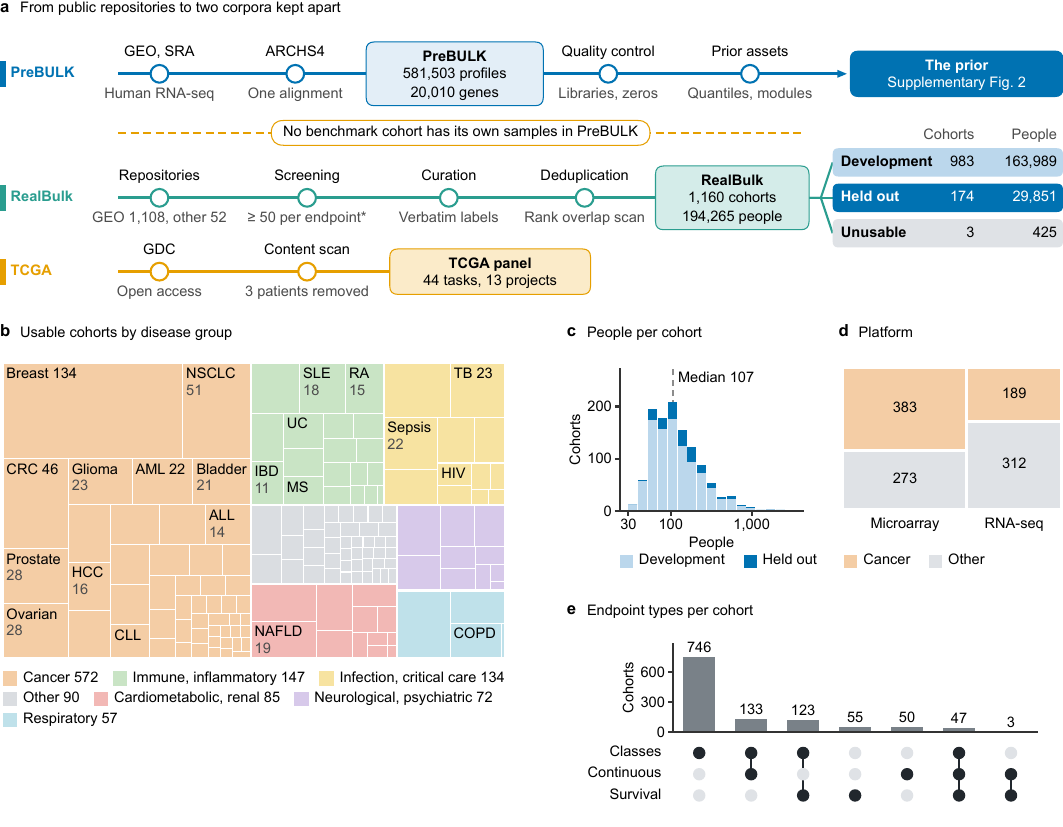}
\suppfigpage{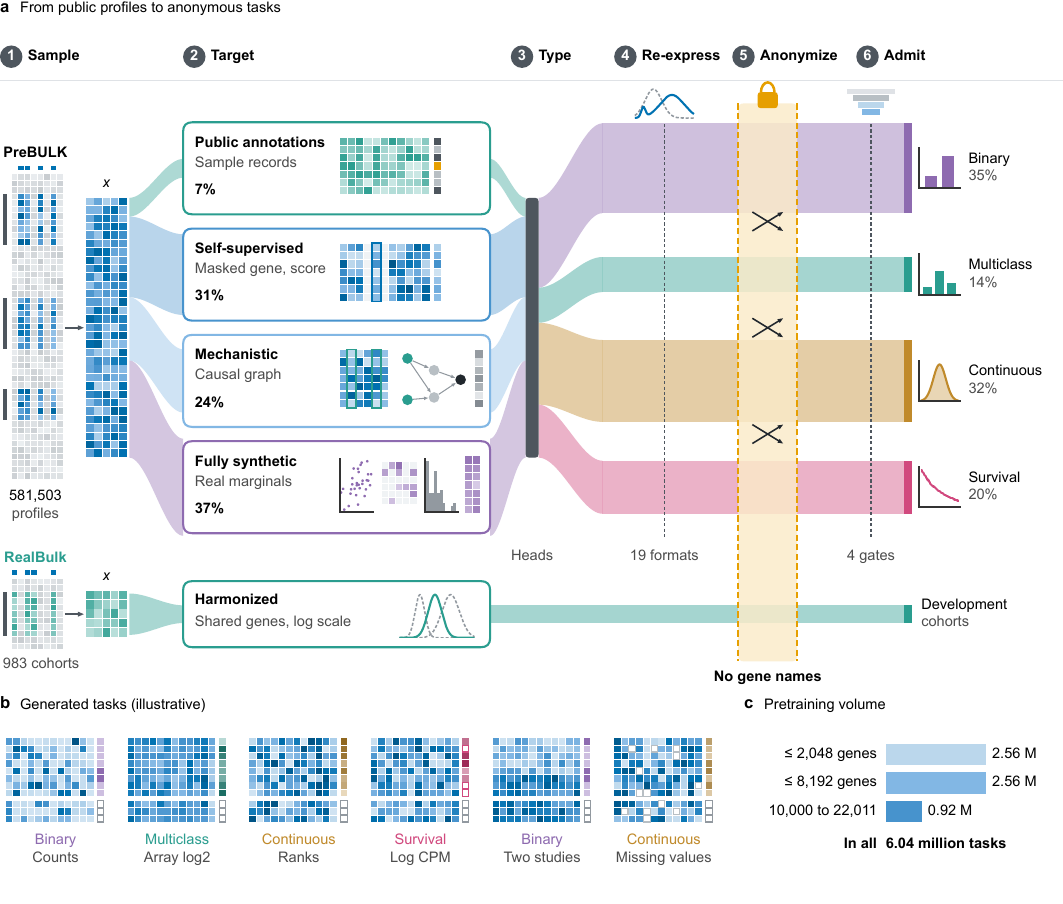}
\suppfigpage{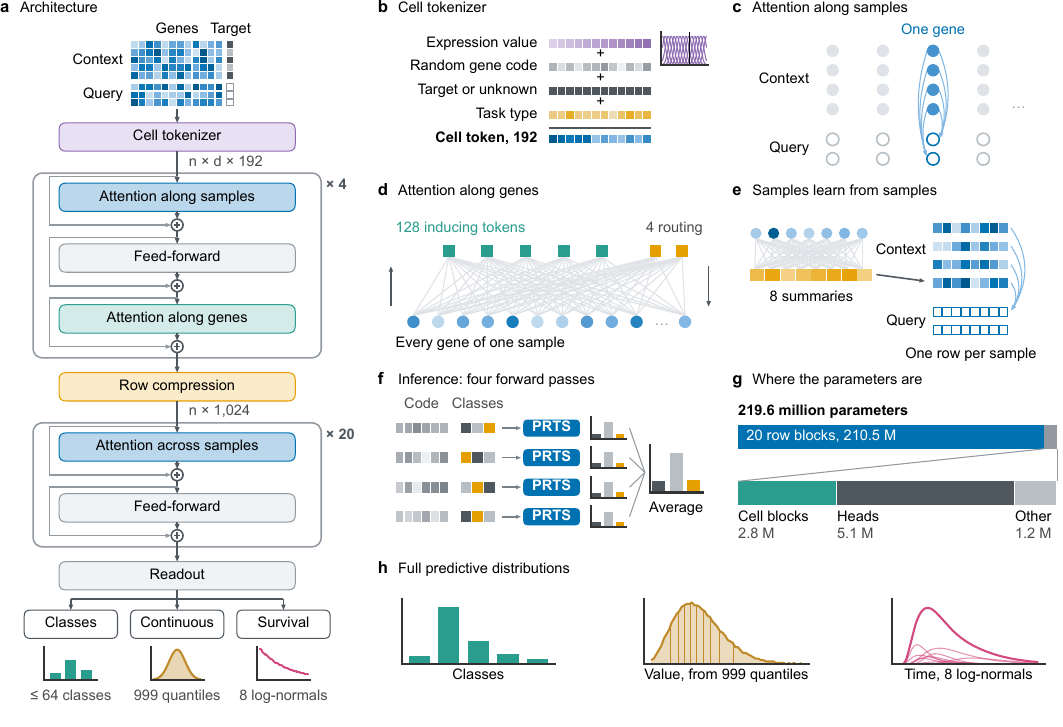}
\suppfigpage{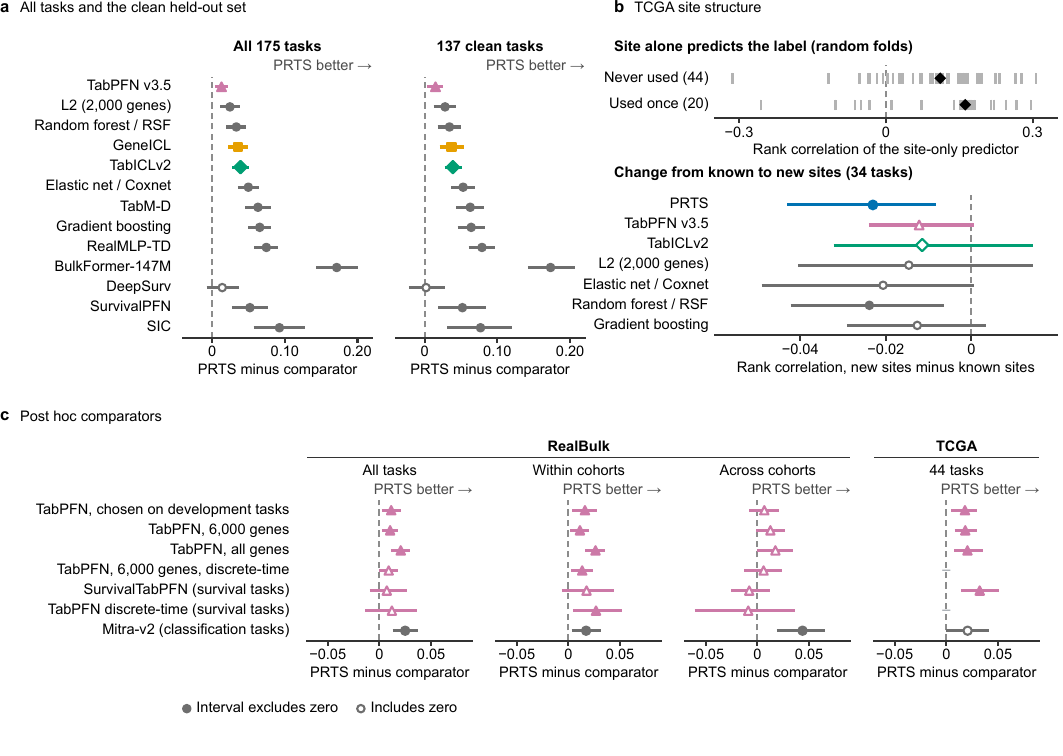}
\suppfigpage{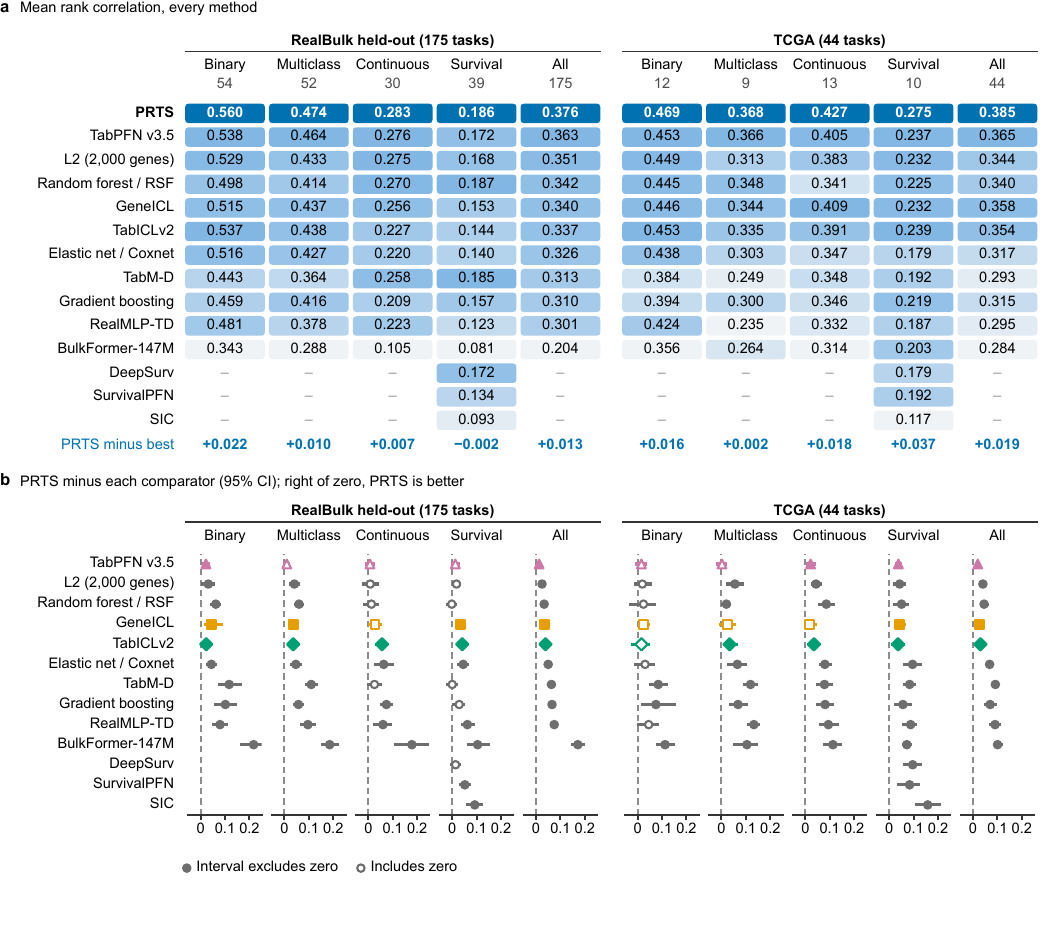}
\suppfigpage{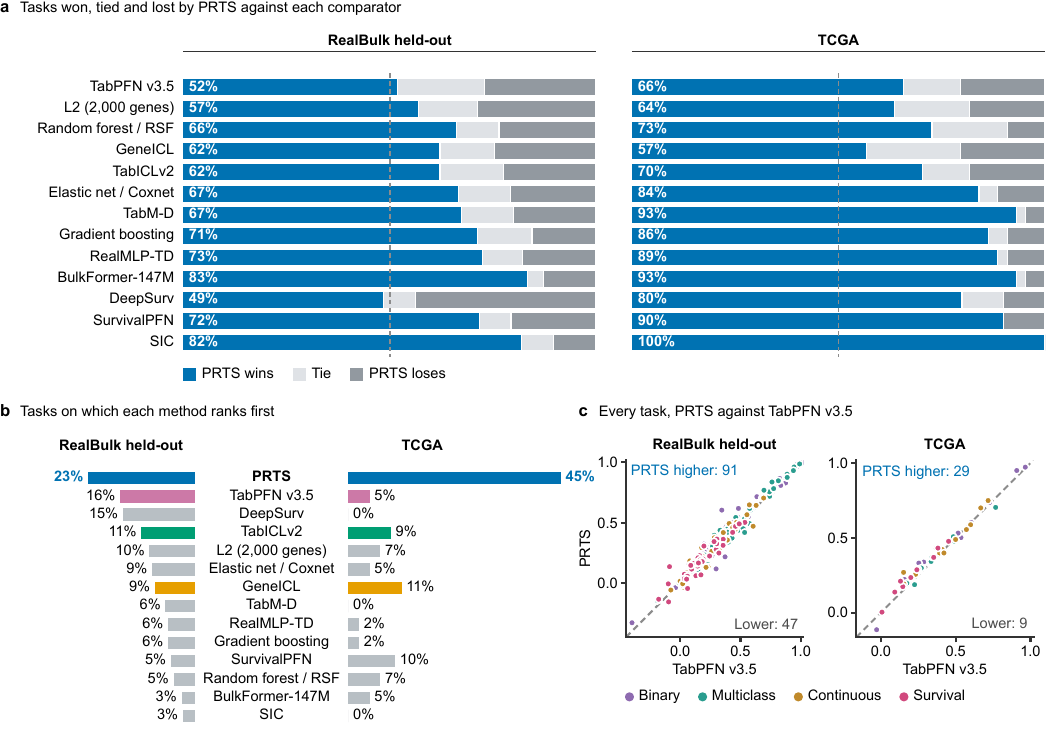}
\suppfigpage{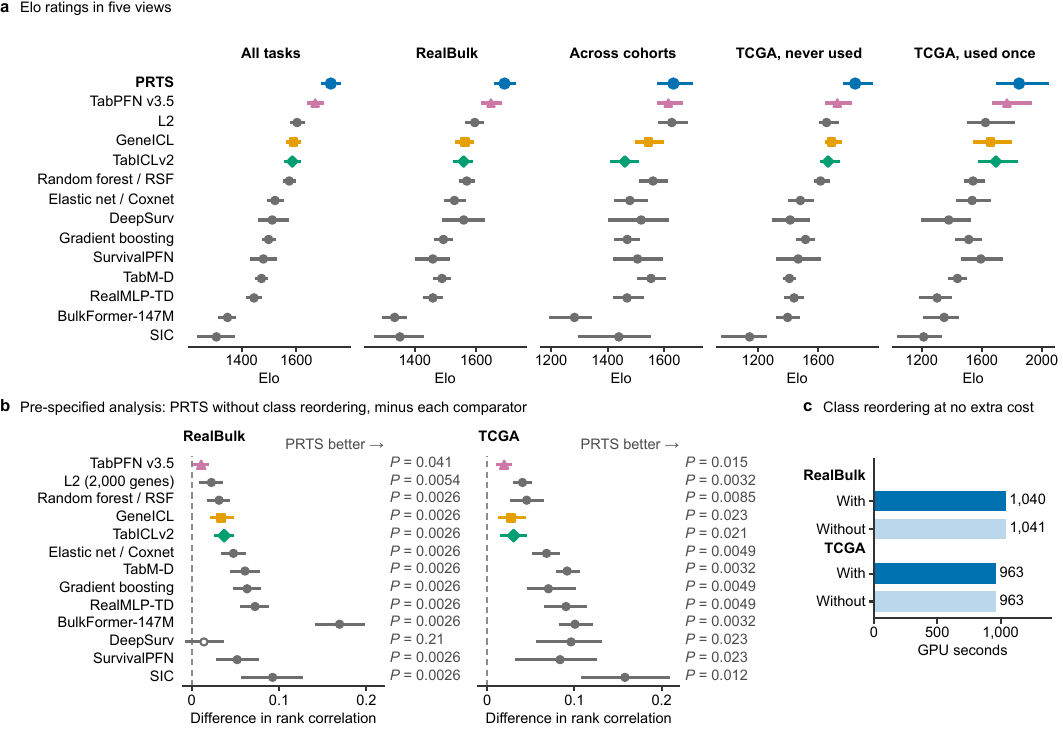}
\suppfigpage{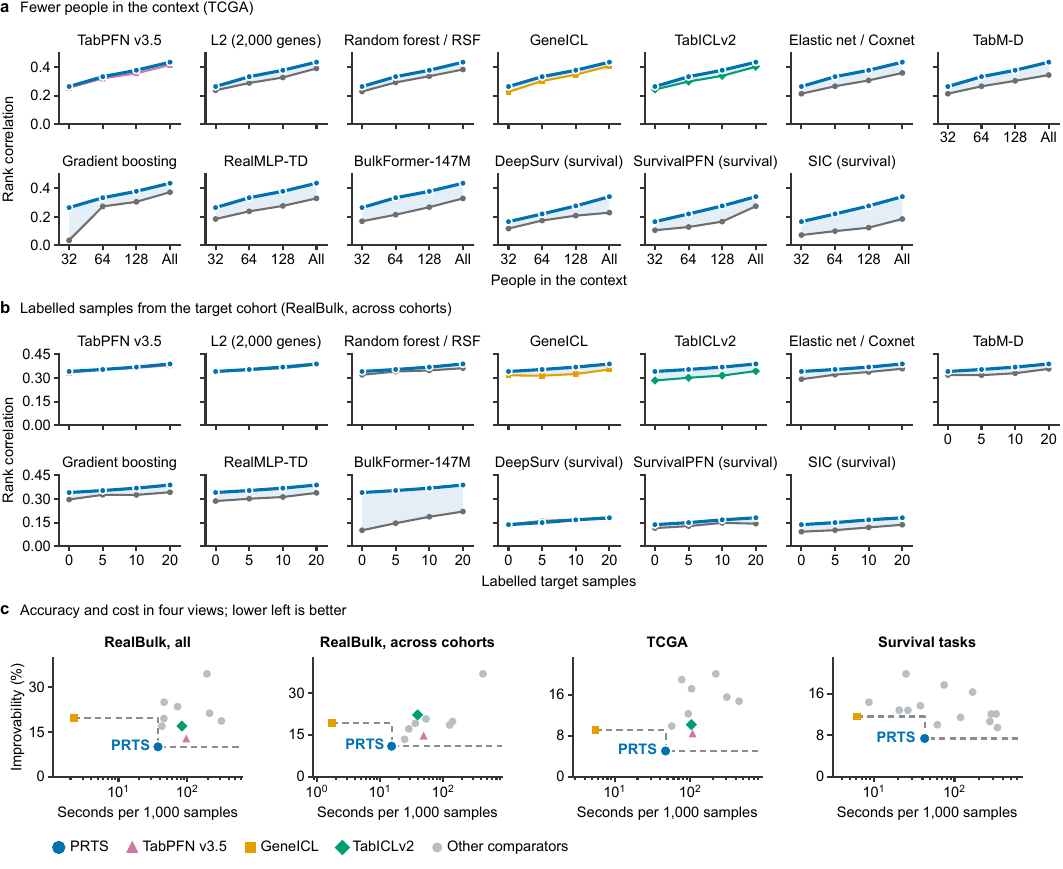}
\suppfigpage{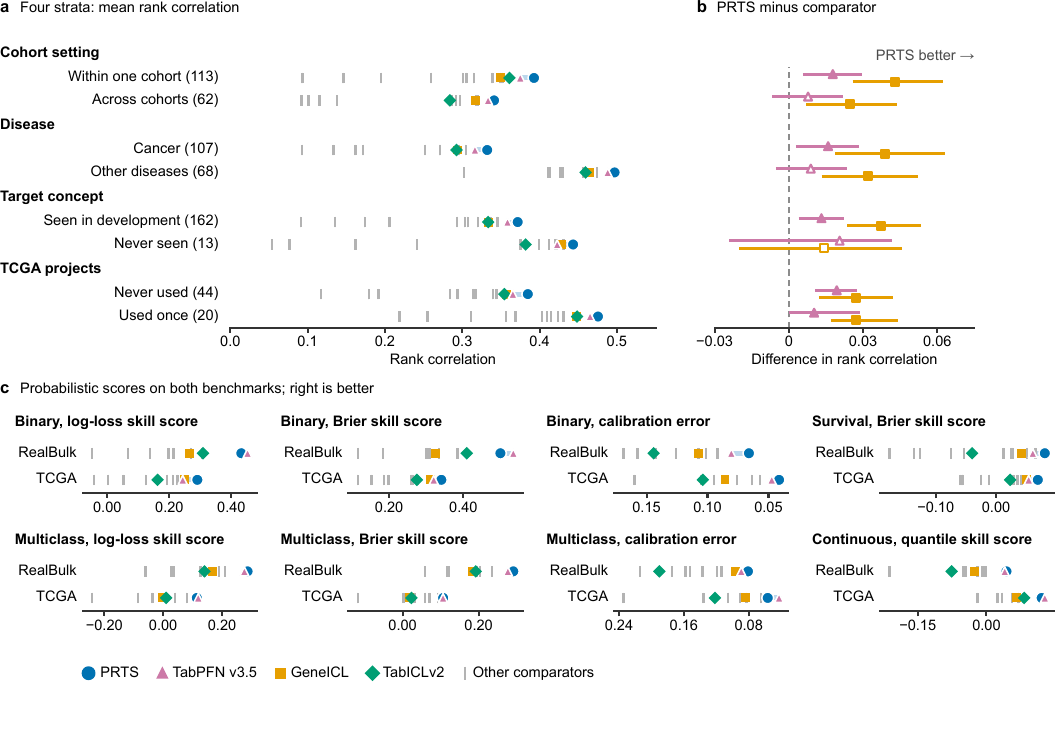}
\restoregeometry

\section*{Supplementary Note 1 \textbar{} The corrected gene ranking and both versions of the affected results}
In an earlier version of this analysis, the comparators that read a fixed number of genes chose them in a way that left
their gene sets arbitrary in cross-cohort tasks. Correcting the choice removed PRTS's cross-cohort lead and reduced its
margin over TabPFN v3.5 on all RealBulk tasks from \val{V1} to \val{B2}; this note describes the defect, its reach and
both versions of every affected result.

In a cross-cohort task the source cohort is the context and the target cohort the query, and each cohort is
standardized on its own samples before any method reads it. Comparators that read a fixed number of genes keep the
2,000 genes of highest variance over the context: the input chosen for TabPFN v3.5 on the development tuning tasks,
and the default of the classical, deep-learning and survival comparators. In the first version, the genes were ranked
after the per-cohort standardization. Every gene then has unit variance over the context, so the ranking fell back on
an arbitrary order and these comparators read an arbitrary 2,000 of the genes the two cohorts share. The same order of
operations affected the Hub's TabPFN v3.5 comparator, whose genes were ranked over reference cohorts that had each been
standardized.

The defect touched the cross-cohort units (the 62 tasks without target labels and the designs with 5, 10 or 20 target
labels) of L2-penalized regression, elastic net, random forests, gradient boosting, $k$-nearest neighbours, TabPFN
v3.5, RealMLP-TD, TabM-D, DeepSurv, DeepHit, SurvivalPFN and SIC. It also touched the Hub's TabPFN v3.5
comparator in both evaluation rounds. It did not touch PRTS or GeneICL (both read every gene), TabICLv2 (the 500
genes most associated with the labels, a choice that does not depend on scale), L2 on every gene, the BulkFormer
embeddings or any within-cohort or TCGA task.

The corrected code ranks the genes by their variance over the context before each cohort is standardized. We repeated
every affected unit with its original settings. No repeated unit fell back to the null model, every repeated run had
the units and query people of the first version, and on the within-cohort units, which the correction leaves alone,
the repeated predictions of L2 and TabPFN v3.5 equal those of the first version. The configurations chosen on the
development tuning tasks did not change. Supplementary Table~4 lists the first and the corrected value of every contrast
that changed; the corrected values replace the first ones throughout the paper.

Without target labels, the margin of PRTS over TabPFN v3.5 went from \val{V2} in the first version to \val{B75} after
the correction, and its margin over L2-penalized regression from \val{V3} to \val{B78}, both intervals now including
zero. With 5, 10 or 20 target labels, margins over TabPFN v3.5 of up to \val{V9} became margins between \val{B81}
and \val{B82}. Over all RealBulk tasks, the lead over TabPFN v3.5 fell from \val{V1} to \val{B2}. In the Hub's
confirmation round, the margin over TabPFN v3.5 given the same references fell from \val{V4} to \val{H2}, and in the
first round from \val{V5} to \val{H30}.

\section*{Supplementary Note 2 \textbar{} Audit of the split between development and held-out cohorts}
The split kept cohorts linked by shared people, accessions, publications or a recorded overlap on one side. An
independent reanalysis of an earlier version of this work found held-out cohorts that nonetheless share patients with
development cohorts, so we rebuilt the link graph from every record of the curation and from the expression data.

\textbf{Sources of links.} Repeat rules and sample exclusion reasons were parsed from each cohort's provenance and
sample tables, and every shared-array file of the curation was read; the original linkage had used none of these. Every
curation note that names a held-out cohort was read and classified by hand. Survival times were matched between every
held-out cohort with a survival endpoint and every cohort with one (6,530 pairs with five or more coinciding times),
up to a unit factor (days, weeks, months or years) and whatever the event coding. Chance counts came from the same
matching with the factor perturbed by 1--3\% or the times shifted by 3--6 months. Expression profiles were compared
between every held-out cohort and every other cohort sharing at least 1,000 genes, on the 3,000 shared genes with the
highest within-cohort variance in both, standardized within each cohort. Samples that are each other's best match were
kept with their margin over the next-best person. A match needs a correlation of at least 0.5 and a margin of at least
0.2: the loosest rule under which at most one in 40,000 of the 180,989 pairs of cohorts from different disease groups,
which cannot share people, has two or more matches. A link needs three or more matches, two at a correlation of at
least 0.8, or one near-identical array.

\textbf{Strength.} A link is strong when it rests on identity (the same array, library, patient code or clinical
record), on two or more recorded people, on a curation note asserting shared people or on an expression link. It is
possible when a record leaves shared people open, such as a note that two series ``must be treated as the same cohort
unless shown disjoint''. A single hedged match to one person, or expression matches below the rule, are weak and form
no link.

\textbf{Results.} The rebuilt graph links \val{K8} cohort pairs, against \val{K7} in the split. It places \val{K1}
held-out cohorts in a component with a development cohort, \val{K2} on strong and \val{K3} on possible evidence;
\val{K15} of them carry benchmark tasks. They include GSE17536, almost wholly contained in the development cohort
GSE14333 (133 coinciding survival times and 169 people matched by expression) and GSE2990, which shares people with
GSE3494, GSE7390 and GSE17705. Others are GSE42822, which holds 59 people of GSE23988; GSE76262, a sputum series of U-BIOPRED participants
also in GSE69683 and GSE76226; and GSE86166, whose 217 arrays recur in GSE199633. Sixteen overlaps that exclusion
reasons recorded had never become links. Two held-out cohorts, GSE38832 and GSE17537, also share people with each
other, which breaks one cross-cohort task. Removing the \val{K4} tasks that touch a linked cohort or cross a missed
source--target link (\val{K5} within one cohort, \val{K6} across cohorts) leaves the clean set of \val{C1} tasks
(Supplementary Tables~1 and 2). No task was added or reselected, and the clean set was fixed before any result on it
was computed.

\textbf{PreBULK.} PreBULK supplies only the profiles from which the prior builds its tasks, never a held-out label.
Matching held-out samples to PreBULK's rows through repaired sample identifiers found \val{K9} held-out RNA-seq cohorts
with samples of their own in PreBULK (GSE210659, GSE224564, GSE231691 and GSE231693), none of which carries a benchmark
task; matching by series accession, where PreBULK records one, found the same four. Through the same people's samples in other series, \val{K10} tasks include people with a sample in PreBULK,
\val{K11} of them in the clean set. An expression scan, calibrated on the 515 held-out samples whose own accession is
in PreBULK, adds candidate matches in \val{K12} tasks; at a false-match rate of about one sample in a thousand, a few of these are expected by
chance.

\textbf{Other findings.} The development cohort GSE14333 had its disease-free survival endpoint stored with the event
indicator inverted and the follow-up times 12-fold too long, because the curation read months as years and the
censoring flag as an event. Of the 127 GSE17536 people whose follow-up time equals one in GSE14333, 124 agree on the event under the
corrected reading and 3 under the stored one. RealBulk's corrected labels use this reading, and no benchmark task
uses GSE14333. Within \val{K14} held-out cohorts, \val{K13} pairs
of near-identical samples carry different person identifiers. Such a pair can sit on both sides of a within-cohort
fold; every method saw the same folds, and we did not measure how much such pairs help each method.

\section*{Supplementary Note 3 \textbar{} Post hoc attempts to improve PRTS}
With the corrected comparators, the margin of PRTS over TabPFN v3.5 on RealBulk fell from \val{V1} to \val{B2}, and PRTS
was level with TabPFN v3.5 across cohorts. We then tried to improve the model on development data only, under rules written down before
each step. No attempt passed its rule, and the released model is unchanged. One candidate remained for a single
evaluation on the held-out sets.

\textbf{Frozen evaluation.} On the development tuning tasks, an equal-weight average of PRTS and TabPFN v3.5 (class
probabilities averaged; continuous medians and survival risks standardized within the task and averaged) exceeded TabPFN
v3.5 alone by \val{X19}. We fixed the evaluation before computing any held-out number: the two methods' existing
held-out predictions, the clean set and TCGA's primary stratum as the primary sets, and the contrasts of the average
with each method. The average led TabPFN v3.5 by \val{X6} (\ci{\val{X7}}{\val{X8}}) on the clean set and by
\val{X12} (\ci{\val{X13}}{\val{X14}}) on TCGA but was level with PRTS (clean set \val{X9}, \ci{\val{X10}}{\val{X11}};
TCGA \val{X15}, \ci{\val{X16}}{\val{X17}}; Supplementary Table~7). Adding TabPFN v3.5 to PRTS
gains nothing on held-out data, so the paper reports PRTS alone.

\clearpage
\section*{Supplementary Tables}

\begingroup\scriptsize
\setlength{\tabcolsep}{3pt}
\begin{longtable}{>{\raggedright\arraybackslash}p{6.4cm}lll>{\raggedright\arraybackslash}p{4.6cm}}
\caption{The benchmark tasks excluded from the clean held-out set.}\label{tab:excluded}\\
\toprule
Task & Kind & Type & Strength & Reason \\
\midrule
\endfirsthead
\multicolumn{5}{l}{\textit{(continued)}}\\
\toprule
Task & Kind & Type & Strength & Reason \\
\midrule
\endhead
\bottomrule
\endfoot
\texttt{breast\_cancer.distant\_recurrence.tumour::}\newline\texttt{GSE22219$\rightarrow$GSE233251} & cross & survival & strong & GSE22219 linked to development; GSE233251 linked to development \\
\texttt{breast\_cancer.grade.tumour::}\newline\texttt{GSE233251$\rightarrow$GSE2990} & cross & multiclass & strong & GSE233251 linked to development; GSE2990 linked to development \\
\texttt{breast\_cancer.grade.tumour::}\newline\texttt{GSE42568$\rightarrow$GSE86166} & cross & multiclass & strong & GSE86166 linked to development \\
\texttt{breast\_cancer.grade.tumour::}\newline\texttt{GSE45255$\rightarrow$GSE86166} & cross & multiclass & strong & GSE45255 linked to development; GSE86166 linked to development \\
\texttt{breast\_cancer.her2\_status.tumour::}\newline\texttt{GSE50948$\rightarrow$GSE45255} & cross & binary & strong & GSE45255 linked to development \\
\texttt{breast\_cancer.pcr\_neoadjuvant.tumour::}\newline\texttt{GSE194040$\rightarrow$GSE42822} & cross & binary & strong & GSE194040 linked to development; GSE42822 linked to development; source GSE194040 and target GSE42822 linked \\
\texttt{breast\_cancer.positive\_nodes.tumour::}\newline\texttt{GSE22219$\rightarrow$GSE76250} & cross & regression & strong & GSE22219 linked to development \\
\texttt{breast\_cancer.pr\_status.tumour::}\newline\texttt{GSE36771$\rightarrow$GSE59590} & cross & binary & strong & GSE59590 linked to development \\
\texttt{breast\_cancer.tumour\_size\_cm.tumour::}\newline\texttt{GSE22219$\rightarrow$GSE19615} & cross & regression & strong & GSE22219 linked to development \\
\texttt{breast\_cancer.tumour\_size\_cm.tumour::}\newline\texttt{GSE45725$\rightarrow$GSE6532\_GPL570} & cross & regression & strong & GSE6532\_GPL570 linked to development \\
\texttt{colorectal\_cancer.stage\_i\_iv.tumour::}\newline\texttt{GSE38832$\rightarrow$GSE17536} & cross & multiclass & strong & GSE38832 linked to development; GSE17536 linked to development; source GSE38832 and target GSE17536 linked \\
\texttt{colorectal\_cancer.stage\_ii\_vs\_iii.tumour::}\newline\texttt{GSE37892$\rightarrow$GSE17537} & cross & binary & strong & GSE37892 linked to development; GSE17537 linked to development \\
\texttt{nsclc.histology.tumour::}\newline\texttt{GSE14814$\rightarrow$GSE19188} & cross & binary & strong & GSE14814 linked to development \\
\texttt{nsclc.os.tumour::}\newline\texttt{GSE157009$\rightarrow$GSE3141} & cross & survival & strong & GSE157009 linked to development \\
\texttt{nsclc.os.tumour::}\newline\texttt{GSE3141$\rightarrow$GSE157010} & cross & survival & strong & GSE157010 linked to development \\
\texttt{nsclc.os.tumour::}\newline\texttt{GSE50081$\rightarrow$GSE68465} & cross & survival & strong & GSE68465 linked to development \\
\texttt{ovarian\_cancer.os.tumour::}\newline\texttt{GSE140082$\rightarrow$GSE26193} & cross & survival & strong & GSE26193 linked to development \\
\texttt{ovarian\_cancer.os.tumour::}\newline\texttt{GSE26193$\rightarrow$GSE51088} & cross & survival & strong & GSE26193 linked to development \\
\texttt{ovarian\_cancer.pfs.tumour::}\newline\texttt{GSE17260$\rightarrow$GSE26193} & cross & survival & strong & GSE26193 linked to development \\
\texttt{prostate\_cancer.psa.tumour::}\newline\texttt{GSE153352$\rightarrow$GSE209954} & cross & regression & strong & GSE209954 linked to development \\
\texttt{prostate\_cancer.psa.tumour::}\newline\texttt{GSE209954$\rightarrow$GSE201805} & cross & regression & strong & GSE209954 linked to development \\
\texttt{psoriasis.diagnosis.skin::}\newline\texttt{GSE212836$\rightarrow$GSE54456} & cross & binary & possible & GSE54456 linked to development \\
\texttt{sle.sledai.whole\_blood::}\newline\texttt{GSE224705$\rightarrow$GSE49454} & cross & regression & strong & GSE224705 linked to development \\
\texttt{GSE135304::cancer\_stage} & within & multiclass & strong & GSE135304 linked to development \\
\texttt{GSE14814::os} & within & survival & strong & GSE14814 linked to development \\
\texttt{GSE150910::lung\_diagnosis} & within & multiclass & possible & GSE150910 linked to development \\
\texttt{GSE157010::overall\_survival} & within & survival & strong & GSE157010 linked to development \\
\texttt{GSE16581::who\_grade} & within & binary & possible & GSE16581 linked to development \\
\texttt{GSE17536::dfs} & within & survival & strong & GSE17536 linked to development \\
\texttt{GSE19151::vte\_status} & within & multiclass & strong & GSE19151 linked to development \\
\texttt{GSE198048::histology} & within & multiclass & strong & GSE198048 linked to development \\
\texttt{GSE209954::pre\_treatment\_psa} & within & regression & strong & GSE209954 linked to development \\
\texttt{GSE235356::progression} & within & binary & strong & GSE235356 linked to development \\
\texttt{GSE36809::diagnosis} & within & binary & possible & GSE36809 linked to development \\
\texttt{GSE38832::ajcc\_pathologic\_stage\_i\_iv} & within & multiclass & strong & GSE38832 linked to development \\
\texttt{GSE38885::immune\_status} & within & binary & strong & GSE38885 linked to development \\
\texttt{GSE68465::os} & within & survival & strong & GSE68465 linked to development \\
\texttt{GSE76262::asthma\_severity} & within & multiclass & strong & GSE76262 linked to development \\
\end{longtable}
\endgroup
{\footnotesize Strong: identity evidence (the same array, library, patient code or clinical record), two or more recorded people, a curation note asserting shared people, or an expression link. Possible: a recorded statement that leaves shared people open. The evidence for every task, hop by hop, is in the released \texttt{excluded\_tasks.csv}.\par}
\medskip

\begingroup\footnotesize
\setlength{\tabcolsep}{3pt}
\begin{longtable}{llll}
\caption{Held-out cohorts that the audit placed in a link component with a development cohort.}\label{tab:linked}\\
\toprule
Held-out cohort & Tasks & Strength & Development cohorts linked directly \\
\midrule
\endfirsthead
\multicolumn{4}{l}{\textit{(continued)}}\\
\toprule
Held-out cohort & Tasks & Strength & Development cohorts linked directly \\
\midrule
\endhead
\bottomrule
\endfoot
GSE135304 & 1 & strong & GSE229705 \\
GSE14814 & 2 & strong & via a held-out cohort \\
GSE157009 & 1 & strong & GSE4573 \\
GSE157010 & 2 & strong & via a held-out cohort \\
GSE17536 & 2 & strong & GSE14333 \\
GSE17537 & 1 & strong & via a held-out cohort \\
GSE19151 & 1 & strong & GSE17156 \\
GSE194040 & 1 & strong & GSE326778 \\
GSE198048 & 1 & strong & GSE252168 \\
GSE209954 & 3 & strong & GSE348708 \\
GSE22219 & 3 & strong & GSE17705, GSE7390 \\
GSE224705 & 1 & strong & GSE121239 \\
GSE233251 & 2 & strong & GSE234114 \\
GSE235356 & 1 & strong & GSE24080 \\
GSE26193 & 3 & strong & GSE19539 \\
GSE2990 & 1 & strong & GSE17705, GSE3494, GSE7390 \\
GSE37892 & 1 & strong & GSE39582 \\
GSE38832 & 2 & strong & via a held-out cohort \\
GSE38885 & 1 & strong & GSE10846 \\
GSE38900 & 0 & strong & GSE103842 \\
GSE42822 & 1 & strong & GSE23988 \\
GSE45255 & 2 & strong & GSE20711 \\
GSE59590 & 1 & strong & GSE55348 \\
GSE6532\_GPL570 & 1 & strong & METABRIC \\
GSE68465 & 2 & strong & GSE68571 \\
GSE76262 & 1 & strong & GSE69683, GSE76226 \\
GSE86166 & 2 & strong & GSE199633 \\
GSE150910 & 1 & possible & via a held-out cohort \\
GSE16581 & 1 & possible & GSE183653 \\
GSE175457 & 0 & possible & GSE47460 \\
GSE231691 & 0 & possible & via a held-out cohort \\
GSE231693 & 0 & possible & via a held-out cohort \\
GSE36809 & 1 & possible & GSE19743 \\
GSE54456 & 1 & possible & GSE13355 \\
\end{longtable}
\endgroup
{\footnotesize Via a held-out cohort: the cohort reaches a development cohort only through another held-out cohort with which it shares people. The full path of every link is in the released \texttt{linked\_cohorts.csv}.\par}
\medskip

\begingroup\scriptsize
\setlength{\tabcolsep}{3pt}
\begin{longtable}{>{\raggedright\arraybackslash}p{2.9cm}>{\raggedright\arraybackslash}p{4.1cm}rrlll}
\caption{Post hoc comparators.}\label{tab:posthoc}\\
\toprule
Tasks & Comparator & \textit{n} & Mean & PRTS minus comparator & \textit{P} & W/T/L \\
\midrule
\endfirsthead
\multicolumn{7}{l}{\textit{(continued)}}\\
\toprule
Tasks & Comparator & \textit{n} & Mean & PRTS minus comparator & \textit{P} & W/T/L \\
\midrule
\endhead
\bottomrule
\endfoot
RealBulk, all tasks & TabPFN, chosen on development tasks & 175 & 0.364 & +0.011 (0.002 to 0.021) & 0.059 & 83/39/53 \\
RealBulk, all tasks & TabPFN, all genes & 175 & 0.355 & +0.021 (0.011 to 0.030) & 0.001 & 100/29/46 \\
RealBulk, all tasks & TabPFN, 6,000 genes & 175 & 0.365 & +0.010 (0.002 to 0.018) & 0.054 & 84/46/45 \\
RealBulk, all tasks & TabPFN, discrete-time & 39 & 0.173 & +0.012 (\textminus{}0.014 to 0.037) & 0.68 & 20/6/13 \\
RealBulk, all tasks & SurvivalTabPFN & 39 & 0.178 & +0.007 (\textminus{}0.009 to 0.026) & 0.68 & 12/10/17 \\
RealBulk, all tasks & Mitra-v2 & 106 & 0.492 & +0.025 (0.013 to 0.037) & 0.002 & 60/26/20 \\
RealBulk, all tasks & TabPFN, 6,000 genes, discrete-time & 175 & 0.367 & +0.009 (0.000 to 0.018) & 0.19 & 83/44/48 \\
RealBulk, within cohorts & TabPFN, chosen on development tasks & 113 & 0.377 & +0.016 (0.004 to 0.028) & 0.061 & 56/24/33 \\
RealBulk, within cohorts & TabPFN, all genes & 113 & 0.367 & +0.026 (0.016 to 0.036) & 0.001 & 63/23/27 \\
RealBulk, within cohorts & TabPFN, 6,000 genes & 113 & 0.382 & +0.011 (0.002 to 0.020) & 0.069 & 52/33/28 \\
RealBulk, within cohorts & TabPFN, discrete-time & 23 & 0.192 & +0.027 (0.004 to 0.052) & 0.069 & 12/6/5 \\
RealBulk, within cohorts & SurvivalTabPFN & 23 & 0.202 & +0.018 (\textminus{}0.006 to 0.044) & 0.18 & 8/6/9 \\
RealBulk, within cohorts & Mitra-v2 & 75 & 0.519 & +0.017 (0.003 to 0.031) & 0.069 & 37/21/17 \\
RealBulk, within cohorts & TabPFN, 6,000 genes, discrete-time & 113 & 0.379 & +0.013 (0.003 to 0.024) & 0.069 & 50/35/28 \\
RealBulk, across cohorts & TabPFN, chosen on development tasks & 62 & 0.334 & +0.007 (\textminus{}0.008 to 0.021) & 1.00 & 27/15/20 \\
RealBulk, across cohorts & TabPFN, all genes & 62 & 0.324 & +0.017 (0.000 to 0.035) & 0.31 & 37/6/19 \\
RealBulk, across cohorts & TabPFN, 6,000 genes & 62 & 0.329 & +0.013 (\textminus{}0.001 to 0.027) & 0.38 & 32/13/17 \\
RealBulk, across cohorts & TabPFN, discrete-time & 16 & 0.146 & \textminus{}0.009 (\textminus{}0.061 to 0.037) & 1.00 & 8/0/8 \\
RealBulk, across cohorts & SurvivalTabPFN & 16 & 0.145 & \textminus{}0.008 (\textminus{}0.025 to 0.012) & 1.00 & 4/4/8 \\
RealBulk, across cohorts & Mitra-v2 & 31 & 0.427 & +0.044 (0.019 to 0.066) & 0.010 & 23/5/3 \\
RealBulk, across cohorts & TabPFN, 6,000 genes, discrete-time & 62 & 0.335 & +0.006 (\textminus{}0.013 to 0.024) & 1.00 & 33/9/20 \\
TCGA, all tasks & TabPFN, chosen on development tasks & 44 & 0.367 & +0.018 (0.005 to 0.029) & 0.050 & 27/7/10 \\
TCGA, all tasks & TabPFN, all genes & 44 & 0.364 & +0.020 (0.008 to 0.036) & 0.047 & 27/10/7 \\
TCGA, all tasks & TabPFN, 6,000 genes & 44 & 0.366 & +0.018 (0.008 to 0.029) & 0.045 & 27/10/7 \\
TCGA, all tasks & SurvivalTabPFN & 10 & 0.243 & +0.032 (0.014 to 0.051) & 0.047 & 7/2/1 \\
TCGA, all tasks & Mitra-v2 & 21 & 0.398 & +0.021 (0.000 to 0.042) & 0.11 & 14/3/4 \\
\end{longtable}
\endgroup
{\footnotesize PRTS minus each post hoc comparator (95\% interval). \textit{P}: Holm-adjusted among the added comparators; RealBulk, link-component bootstrap; TCGA, exact project-level sign-flip. W/T/L: tasks PRTS wins, ties (within 0.01) and loses. TabPFN, chosen on development tasks: 2,000 genes with SurvivalTabPFN on survival tasks; TabPFN, 6,000 genes, discrete-time: 6,000 genes with the discrete-time extension on survival tasks. Both combined configurations were assembled from the other runs without new predictions. SurvivalTabPFN and the discrete-time extension: survival tasks only; Mitra-v2: classification tasks only. The discrete-time extension's TCGA run was stopped unread.\par}
\medskip

\begingroup\scriptsize
\setlength{\tabcolsep}{3pt}
\begin{longtable}{llllll}
\caption{Both versions of every contrast changed by the correction of the comparators' gene ranking.}\label{tab:versions}\\
\toprule
Comparator & Design & First version & \textit{P} & Corrected & \textit{P} \\
\midrule
\endfirsthead
\multicolumn{6}{l}{\textit{(continued)}}\\
\toprule
Comparator & Design & First version & \textit{P} & Corrected & \textit{P} \\
\midrule
\endhead
\bottomrule
\endfoot
\midrule\multicolumn{6}{l}{\textit{RealBulk, all 175 tasks}} \\
L2 linear & All tasks & +0.042 (0.030 to 0.055) & 0.003 & +0.025 (0.011 to 0.038) & 0.003 \\
Elastic net / Coxnet & All tasks & +0.065 (0.050 to 0.081) & 0.003 & +0.050 (0.035 to 0.064) & 0.003 \\
Random forest / RSF & All tasks & +0.051 (0.037 to 0.065) & 0.003 & +0.033 (0.020 to 0.046) & 0.003 \\
Gradient boosting & All tasks & +0.081 (0.062 to 0.099) & 0.003 & +0.065 (0.050 to 0.081) & 0.003 \\
kNN & All tasks & +0.082 (0.064 to 0.101) & -- & +0.064 (0.045 to 0.083) & -- \\
TabPFN v3.5 & All tasks & +0.022 (0.012 to 0.032) & 0.003 & +0.013 (0.004 to 0.022) & 0.016 \\
RealMLP-TD & All tasks & +0.080 (0.063 to 0.095) & 0.003 & +0.075 (0.058 to 0.091) & 0.003 \\
TabM-D & All tasks & +0.080 (0.064 to 0.097) & 0.003 & +0.063 (0.046 to 0.080) & 0.003 \\
DeepSurv & All tasks & +0.039 (0.018 to 0.062) & 0.003 & +0.014 (\textminus{}0.008 to 0.037) & 0.21 \\
DeepHit & All tasks & +0.075 (0.048 to 0.101) & -- & +0.048 (0.010 to 0.084) & -- \\
SurvivalPFN & All tasks & +0.061 (0.034 to 0.090) & 0.003 & +0.052 (0.027 to 0.077) & 0.003 \\
SIC & All tasks & +0.121 (0.090 to 0.154) & 0.003 & +0.093 (0.057 to 0.128) & 0.003 \\
\midrule\multicolumn{6}{l}{\textit{RealBulk across cohorts (62 tasks) and within cohorts (113)}} \\
L2 linear & 0 target labels & +0.050 (0.029 to 0.071) & 0.003 & +0.001 (\textminus{}0.025 to 0.026) & 1.00 \\
Elastic net / Coxnet & 0 target labels & +0.089 (0.061 to 0.118) & 0.003 & +0.049 (0.027 to 0.073) & 0.003 \\
Random forest / RSF & 0 target labels & +0.067 (0.041 to 0.093) & 0.003 & +0.021 (\textminus{}0.001 to 0.042) & 0.45 \\
Gradient boosting & 0 target labels & +0.086 (0.057 to 0.115) & 0.003 & +0.044 (0.025 to 0.063) & 0.003 \\
kNN & 0 target labels & +0.084 (0.054 to 0.115) & -- & +0.031 (\textminus{}0.004 to 0.063) & -- \\
TabPFN v3.5 & 0 target labels & +0.033 (0.014 to 0.052) & 0.003 & +0.008 (\textminus{}0.007 to 0.022) & 0.94 \\
TabICLv2 & 0 target labels & +0.057 (0.035 to 0.080) & 0.003 & +0.057 (0.035 to 0.080) & 0.003 \\
RealMLP-TD & 0 target labels & +0.067 (0.038 to 0.096) & 0.003 & +0.054 (0.023 to 0.084) & 0.003 \\
TabM-D & 0 target labels & +0.069 (0.042 to 0.097) & 0.003 & +0.023 (\textminus{}0.007 to 0.053) & 0.79 \\
DeepSurv & 0 target labels & +0.061 (0.029 to 0.099) & 0.003 & \textminus{}0.001 (\textminus{}0.036 to 0.037) & 1.00 \\
DeepHit & 0 target labels & +0.060 (0.013 to 0.107) & -- & \textminus{}0.006 (\textminus{}0.074 to 0.063) & -- \\
SurvivalPFN & 0 target labels & +0.044 (0.000 to 0.094) & 0.048 & +0.022 (\textminus{}0.013 to 0.056) & 0.89 \\
SIC & 0 target labels & +0.115 (0.054 to 0.181) & 0.003 & +0.045 (\textminus{}0.014 to 0.110) & 0.79 \\
GeneICL & 0 target labels & +0.025 (0.006 to 0.044) & 0.018 & +0.025 (0.006 to 0.044) & 0.070 \\
L2 linear & 5 target labels & +0.053 (0.038 to 0.069) & 0.003 & +0.003 (\textminus{}0.015 to 0.020) & 1.00 \\
Elastic net / Coxnet & 5 target labels & +0.072 (0.052 to 0.092) & 0.003 & +0.033 (0.018 to 0.048) & 0.003 \\
Random forest / RSF & 5 target labels & +0.070 (0.044 to 0.097) & 0.003 & +0.013 (\textminus{}0.008 to 0.036) & 1.00 \\
Gradient boosting & 5 target labels & +0.084 (0.061 to 0.109) & 0.003 & +0.027 (0.011 to 0.043) & 0.011 \\
kNN & 5 target labels & +0.070 (0.047 to 0.095) & -- & +0.033 (0.008 to 0.059) & -- \\
TabPFN v3.5 & 5 target labels & +0.037 (0.022 to 0.052) & 0.003 & +0.000 (\textminus{}0.012 to 0.012) & 1.00 \\
TabICLv2 & 5 target labels & +0.053 (0.038 to 0.068) & 0.003 & +0.053 (0.038 to 0.068) & 0.003 \\
RealMLP-TD & 5 target labels & +0.090 (0.067 to 0.114) & 0.003 & +0.052 (0.030 to 0.074) & 0.003 \\
TabM-D & 5 target labels & +0.082 (0.058 to 0.110) & 0.003 & +0.036 (0.015 to 0.060) & 0.011 \\
DeepSurv & 5 target labels & +0.039 (0.003 to 0.074) & 0.037 & \textminus{}0.011 (\textminus{}0.043 to 0.023) & 1.00 \\
DeepHit & 5 target labels & +0.086 (0.044 to 0.136) & -- & +0.027 (\textminus{}0.017 to 0.075) & -- \\
SurvivalPFN & 5 target labels & +0.042 (0.010 to 0.080) & 0.023 & +0.022 (\textminus{}0.019 to 0.057) & 1.00 \\
SIC & 5 target labels & +0.116 (0.073 to 0.163) & 0.003 & +0.048 (\textminus{}0.001 to 0.103) & 0.34 \\
GeneICL & 5 target labels & +0.040 (0.024 to 0.056) & 0.003 & +0.040 (0.024 to 0.056) & 0.003 \\
L2 linear & 10 target labels & +0.049 (0.033 to 0.064) & 0.003 & +0.007 (\textminus{}0.012 to 0.026) & 1.00 \\
Elastic net / Coxnet & 10 target labels & +0.082 (0.062 to 0.104) & 0.003 & +0.031 (0.011 to 0.051) & 0.021 \\
Random forest / RSF & 10 target labels & +0.071 (0.044 to 0.100) & 0.003 & +0.021 (0.000 to 0.045) & 0.30 \\
Gradient boosting & 10 target labels & +0.084 (0.061 to 0.108) & 0.003 & +0.043 (0.027 to 0.059) & 0.003 \\
kNN & 10 target labels & +0.084 (0.059 to 0.110) & -- & +0.043 (0.020 to 0.067) & -- \\
TabPFN v3.5 & 10 target labels & +0.041 (0.027 to 0.055) & 0.003 & +0.004 (\textminus{}0.008 to 0.017) & 1.00 \\
TabICLv2 & 10 target labels & +0.054 (0.039 to 0.068) & 0.003 & +0.054 (0.039 to 0.068) & 0.003 \\
RealMLP-TD & 10 target labels & +0.088 (0.066 to 0.109) & 0.003 & +0.056 (0.035 to 0.078) & 0.003 \\
TabM-D & 10 target labels & +0.072 (0.051 to 0.093) & 0.003 & +0.039 (0.017 to 0.062) & 0.006 \\
DeepSurv & 10 target labels & +0.070 (0.040 to 0.103) & 0.003 & \textminus{}0.002 (\textminus{}0.044 to 0.045) & 1.00 \\
DeepHit & 10 target labels & +0.090 (0.059 to 0.125) & -- & +0.069 (0.026 to 0.120) & -- \\
SurvivalPFN & 10 target labels & +0.037 (0.016 to 0.062) & 0.003 & +0.019 (\textminus{}0.014 to 0.051) & 1.00 \\
SIC & 10 target labels & +0.114 (0.066 to 0.165) & 0.003 & +0.048 (\textminus{}0.010 to 0.110) & 0.52 \\
GeneICL & 10 target labels & +0.044 (0.026 to 0.063) & 0.003 & +0.044 (0.026 to 0.063) & 0.003 \\
L2 linear & 20 target labels & +0.046 (0.028 to 0.064) & 0.003 & +0.004 (\textminus{}0.012 to 0.021) & 1.00 \\
Elastic net / Coxnet & 20 target labels & +0.073 (0.052 to 0.094) & 0.003 & +0.030 (0.011 to 0.050) & 0.018 \\
Random forest / RSF & 20 target labels & +0.088 (0.058 to 0.121) & 0.003 & +0.027 (0.004 to 0.053) & 0.11 \\
Gradient boosting & 20 target labels & +0.100 (0.075 to 0.126) & 0.003 & +0.046 (0.026 to 0.066) & 0.003 \\
kNN & 20 target labels & +0.101 (0.076 to 0.127) & -- & +0.053 (0.026 to 0.081) & -- \\
TabPFN v3.5 & 20 target labels & +0.040 (0.025 to 0.055) & 0.003 & +0.008 (\textminus{}0.005 to 0.022) & 0.73 \\
TabICLv2 & 20 target labels & +0.045 (0.031 to 0.059) & 0.003 & +0.045 (0.031 to 0.059) & 0.003 \\
RealMLP-TD & 20 target labels & +0.115 (0.093 to 0.138) & 0.003 & +0.050 (0.027 to 0.073) & 0.003 \\
TabM-D & 20 target labels & +0.064 (0.042 to 0.087) & 0.003 & +0.031 (0.006 to 0.057) & 0.10 \\
DeepSurv & 20 target labels & +0.065 (0.024 to 0.100) & 0.011 & +0.006 (\textminus{}0.040 to 0.047) & 1.00 \\
DeepHit & 20 target labels & +0.083 (0.030 to 0.127) & -- & +0.039 (\textminus{}0.021 to 0.099) & -- \\
SurvivalPFN & 20 target labels & +0.047 (0.016 to 0.076) & 0.011 & +0.038 (\textminus{}0.002 to 0.069) & 0.31 \\
SIC & 20 target labels & +0.083 (0.029 to 0.136) & 0.011 & +0.044 (\textminus{}0.021 to 0.115) & 0.73 \\
GeneICL & 20 target labels & +0.035 (0.021 to 0.052) & 0.003 & +0.035 (0.021 to 0.052) & 0.003 \\
L2 linear & Within cohorts & +0.041 (0.025 to 0.057) & 0.003 & +0.041 (0.025 to 0.057) & 0.003 \\
Random forest / RSF & Within cohorts & +0.043 (0.027 to 0.060) & 0.003 & +0.043 (0.027 to 0.060) & 0.003 \\
TabPFN v3.5 & Within cohorts & +0.018 (0.005 to 0.029) & 0.010 & +0.018 (0.005 to 0.029) & 0.010 \\
TabICLv2 & Within cohorts & +0.032 (0.018 to 0.044) & 0.003 & +0.032 (0.018 to 0.044) & 0.003 \\
GeneICL & Within cohorts & +0.043 (0.025 to 0.062) & 0.003 & +0.043 (0.025 to 0.062) & 0.003 \\
\midrule\multicolumn{6}{l}{\textit{Hub minus comparator (task families weighted equally)}} \\
TabPFN v3.5 (same references) & First round & +0.043 (0.021 to 0.069) & -- & +0.022 (0.002 to 0.044) & -- \\
TabPFN v3.5 (same references) & Confirmation & +0.028 (0.013 to 0.045) & -- & +0.013 (\textminus{}0.001 to 0.027) & -- \\
\end{longtable}
\endgroup
{\footnotesize PRTS minus each comparator in mean rank correlation (95\% cluster-bootstrap interval) and the Holm-adjusted \textit{P} value, in the first version and after the correction of the comparators' gene ranking. Only cross-cohort units of comparators that read a variance-ranked gene set changed; GeneICL, TabICLv2, BulkFormer-147M and L2 on every gene were unaffected, as were all within-cohort and TCGA results. The across- and within-cohort section shows every changed contrast and, for contrast, the main comparators. Hub rows: the first round and the confirmation. Elo ratings, first version to corrected: RealBulk: PRTS 1716 to 1691, TabPFN v3.5 1639 to 1647; RealBulk across cohorts: PRTS 1696 to 1633, TabPFN v3.5 1598 to 1614; all tasks: PRTS 1746 to 1726, TabPFN v3.5 1662 to 1668.\par}
\medskip

\begingroup\scriptsize
\setlength{\tabcolsep}{3pt}
\begin{longtable}{>{\raggedright\arraybackslash}p{3.1cm}rrr>{\raggedright\arraybackslash}p{2.3cm}>{\raggedright\arraybackslash}p{2.3cm}>{\raggedright\arraybackslash}p{2.3cm}l}
\caption{TCGA with random and site-grouped folds.}\label{tab:sites}\\
\toprule
Method & \textit{n} & Known sites & New sites & Change & PRTS minus, known sites & PRTS minus, new sites & \textit{P} \\
\midrule
\endfirsthead
\multicolumn{8}{l}{\textit{(continued)}}\\
\toprule
Method & \textit{n} & Known sites & New sites & Change & PRTS minus, known sites & PRTS minus, new sites & \textit{P} \\
\midrule
\endhead
\bottomrule
\endfoot
PRTS & 34 & 0.401 & 0.378 & \textminus{}0.023\newline(\textminus{}0.043 to \textminus{}0.008) & -- & -- & -- \\
TabPFN v3.5 & 34 & 0.382 & 0.370 & \textminus{}0.012\newline(\textminus{}0.024 to 0.001) & +0.018\newline(0.008 to 0.029) & +0.007\newline(\textminus{}0.007 to 0.021) & 0.39 \\
TabICLv2 & 34 & 0.369 & 0.358 & \textminus{}0.012\newline(\textminus{}0.032 to 0.014) & +0.032\newline(0.018 to 0.048) & +0.020\newline(0.000 to 0.034) & 0.070 \\
L2 linear & 34 & 0.364 & 0.349 & \textminus{}0.015\newline(\textminus{}0.040 to 0.014) & +0.037\newline(0.029 to 0.045) & +0.029\newline(0.008 to 0.046) & 0.039 \\
Elastic net / Coxnet & 34 & 0.329 & 0.308 & \textminus{}0.021\newline(\textminus{}0.049 to 0.000) & +0.072\newline(0.056 to 0.088) & +0.069\newline(0.052 to 0.090) & 0.004 \\
Random forest / RSF & 34 & 0.358 & 0.334 & \textminus{}0.024\newline(\textminus{}0.042 to \textminus{}0.006) & +0.042\newline(0.022 to 0.062) & +0.043\newline(0.015 to 0.069) & 0.031 \\
Gradient boosting & 34 & 0.338 & 0.325 & \textminus{}0.013\newline(\textminus{}0.029 to 0.003) & +0.063\newline(0.044 to 0.085) & +0.052\newline(0.038 to 0.065) & 0.004 \\
DeepSurv & 8 & 0.173 & 0.170 & \textminus{}0.003\newline(\textminus{}0.052 to 0.049) & +0.103\newline(0.059 to 0.139) & +0.071\newline(0.030 to 0.115) & 0.023 \\
DeepHit & 8 & 0.170 & 0.133 & \textminus{}0.037\newline(\textminus{}0.077 to \textminus{}0.005) & +0.106\newline(0.052 to 0.160) & +0.108\newline(0.056 to 0.159) & 0.016 \\
SurvivalPFN & 8 & 0.191 & 0.159 & \textminus{}0.031\newline(\textminus{}0.082 to 0.018) & +0.086\newline(0.021 to 0.136) & +0.082\newline(0.022 to 0.148) & 0.055 \\
SIC & 8 & 0.119 & 0.098 & \textminus{}0.021\newline(\textminus{}0.072 to 0.030) & +0.158\newline(0.095 to 0.222) & +0.144\newline(0.075 to 0.221) & 0.008 \\
\end{longtable}
\endgroup
{\footnotesize TCGA primary-stratum tasks where site-grouped folds are feasible (at least eight sites, no site above 30\% of patients). Mean rank correlation with random folds (known sites) and site-grouped folds (new sites), the change (bootstrap over projects), and PRTS minus each method in both designs; \textit{P}: exact project-level sign-flip test of the site-grouped contrast. Survival-only methods: survival tasks.\par}
\medskip

\begingroup\scriptsize
\setlength{\tabcolsep}{3pt}
\begin{longtable}{ll>{\raggedright\arraybackslash}p{5.6cm}>{\raggedright\arraybackslash}p{2.9cm}>{\raggedright\arraybackslash}p{2.9cm}}
\caption{The four applications on all held-out items and on items from clean cohorts.}\label{tab:tools}\\
\toprule
Application & Round & Contrast & All items & Clean items \\
\midrule
\endfirsthead
\multicolumn{5}{l}{\textit{(continued)}}\\
\toprule
Application & Round & Contrast & All items & Clean items \\
\midrule
\endhead
\bottomrule
\endfoot
Hub & first round & Hub minus pooled L2 (penalty tuned on the references) & 45: +0.015\newline(\textminus{}0.003 to 0.031) & 39: +0.010\newline(\textminus{}0.009 to 0.028) \\
Hub & first round & Hub minus SingleR-style correlation & 34: +0.039\newline(0.003 to 0.082) & 30: +0.043\newline(0.001 to 0.089) \\
Hub & first round & Hub minus GeneICL (same references) & 45: +0.017\newline(\textminus{}0.001 to 0.035) & 39: +0.019\newline(\textminus{}0.001 to 0.041) \\
Hub & first round & Hub minus TabPFN v3.5 (same references) & 45: +0.022\newline(0.002 to 0.044) & 39: +0.014\newline(\textminus{}0.006 to 0.037) \\
Hub & first round & Hub minus a single reference & 45: +0.015\newline(\textminus{}0.010 to 0.041) & 39: +0.012\newline(\textminus{}0.015 to 0.043) \\
Hub & confirmation & Hub minus pooled L2 (penalty tuned on the references) & 73: +0.011\newline(\textminus{}0.006 to 0.028) & 47: +0.026\newline(0.008 to 0.047) \\
Hub & confirmation & Hub minus SingleR-style correlation & 48: +0.059\newline(0.038 to 0.081) & 34: +0.063\newline(0.039 to 0.087) \\
Hub & confirmation & Hub minus GeneICL (same references) & 73: +0.015\newline(0.001 to 0.030) & 47: +0.016\newline(0.002 to 0.031) \\
Hub & confirmation & Hub minus TabPFN v3.5 (same references) & 73: +0.013\newline(\textminus{}0.001 to 0.027) & 47: +0.020\newline(0.003 to 0.037) \\
Hub & confirmation & Hub minus a single reference & 73: +0.035\newline(0.017 to 0.054) & 47: +0.038\newline(0.015 to 0.061) \\
Planner & confirmation & Planner minus least squares, mean absolute error & 64: \textminus{}0.080\newline(\textminus{}0.108 to \textminus{}0.057) & 56: \textminus{}0.084\newline(\textminus{}0.115 to \textminus{}0.060) \\
Audit & first round & Audit minus random forest, average precision & 47: +0.045\newline(0.022 to 0.070) & 41: +0.045\newline(0.020 to 0.071) \\
Explain & first round & mean rank, PRTS & 21: 4.10 & 18: 4.19 \\
Explain & first round & mean rank, moderated t & 21: 7.55 & 18: 7.64 \\
Explain & first round & mean rank, preranked GSEA on moderated t & 21: 7.43 & 18: 7.78 \\
Explain & first round & mean rank, L2 coefficients & 21: 10.19 & 18: 11.00 \\
Explain & first round & mean rank, LightGBM SHAP values & 21: 6.07 & 18: 6.08 \\
\end{longtable}
\endgroup
{\footnotesize Each application's headline contrast on all held-out items and on the items whose cohort shares no people with a development cohort under the rebuilt link graph; number of items, then the estimate (95\% interval from the report's own cluster bootstrap). Explain module ranks are descriptive.\par}
\medskip

\begingroup\scriptsize
\setlength{\tabcolsep}{3pt}
\begin{longtable}{>{\raggedright\arraybackslash}p{7.2cm}>{\raggedright\arraybackslash}p{3.0cm}>{\raggedright\arraybackslash}p{4.6cm}}
\caption{Post hoc evaluation of an average of PRTS and TabPFN v3.5.}\label{tab:attempts}\\
\toprule
Contrast & Difference (95\% CI) & Test \\
\midrule
\endfirsthead
\multicolumn{3}{l}{\textit{(continued)}}\\
\toprule
Contrast & Difference (95\% CI) & Test \\
\midrule
\endhead
\bottomrule
\endfoot
RealBulk, all tasks: ensemble minus TabPFN v3.5 & +0.013\newline(0.008 to 0.019) & \textit{P} $<$ 0.001; W/T/L 94/55/26 \\
RealBulk, all tasks: ensemble minus PRTS & +0.000\newline(\textminus{}0.004 to 0.005) & \textit{P} = 0.85; W/T/L 48/72/55 \\
RealBulk, clean set: ensemble minus TabPFN v3.5 & +0.015\newline(0.009 to 0.021) & \textit{P} $<$ 0.001; W/T/L 77/40/20 \\
RealBulk, clean set: ensemble minus PRTS & +0.000\newline(\textminus{}0.005 to 0.006) & \textit{P} = 0.88; W/T/L 38/54/45 \\
TCGA, primary stratum: ensemble minus TabPFN v3.5 & +0.017\newline(0.012 to 0.022) & \textit{P} $<$ 0.001; W/T/L 30/10/4 \\
TCGA, primary stratum: ensemble minus PRTS & \textminus{}0.002\newline(\textminus{}0.007 to 0.002) & \textit{P} = 0.30; W/T/L 10/21/13 \\
TCGA, projects used once to compare checkpoints: ensemble minus TabPFN v3.5 & +0.012\newline(0.007 to 0.021) & \textit{P} $<$ 0.001; W/T/L 11/9/0 \\
TCGA, projects used once to compare checkpoints: ensemble minus PRTS & +0.002\newline(\textminus{}0.008 to 0.008) & \textit{P} = 0.91; W/T/L 3/12/5 \\
\end{longtable}
\endgroup
{\footnotesize An equal-weight average of PRTS and TabPFN v3.5, weight chosen on the development tuning tasks, evaluated once under a protocol fixed before any held-out number was computed; differences with 95\% intervals, Holm-adjusted \textit{P} values, and wins, ties and losses (W/T/L).\par}
\medskip

\end{document}